\documentclass{article}
\usepackage{amssymb,amsmath,amsthm,graphicx,xcolor} 
\usepackage{graphics,color}
\usepackage{subfigure}
\usepackage{latexsym}
\usepackage{bm}
\usepackage[hang,flushmargin]{footmisc}
\usepackage{epsfig}
\usepackage{textcomp}
\usepackage[margin=2cm]{geometry}
\usepackage{wasysym}
\usepackage{hyperref}
\usepackage[font=small,labelfont=bf]{caption}

\newtheorem{prop}{Proposition}

\newcommand{\q}[0]{\mathfrak{q}}
\newcommand{\e}[0]{\mathfrak{e}}
\newcommand{\m}[0]{\mathfrak{m}}

\title{Formation of extremal Reissner-Nordstr\"om black holes:\\ insights from numerics}
\author{Maxime~Gadioux, Harvey.~S.~Reall and Jorge~E.~Santos \\ {\footnotesize Department of Applied Mathematics and Theoretical Physics, University of Cambridge} \\ {\footnotesize Wilberforce Road, Cambridge CB3 0WA, United Kingdom}\\ {\footnotesize mjhg2@cam.ac.uk, hsr1000@cam.ac.uk, jss55@cam.ac.uk}}

\date{\today}

\begin{document}

\maketitle

\begin{abstract}
    \noindent An extremal Reissner-Nordstr\"om black hole can form in finite time in the gravitational collapse of a massless charged scalar field. The proof of this is based on the method of characteristic gluing, which involves making an Ansatz for the scalar field at the horizon. We perform a numerical investigation of the characteristic gluing procedure for several different Ans\"atze. In each case, gluing is possible only if the final black hole mass is large enough. We find that the minimum required mass varies significantly for different Ans\"atze. We also consider the effect of including a mass term for the scalar field. In this case, for each Ansatz we determine the maximum mass-to-charge ratio for the scalar field such that gluing is possible. Analogous results are obtained for a non-zero cosmological constant.
\end{abstract}

\section{Introduction}

The third law of black hole mechanics states that it is not possible for a non-extremal black hole to become extremal in finite time in any (classical) process, no matter how idealised, involving physically reasonable matter. This was first conjectured by Bardeen, Carter and Hawking \cite{Bardeen:1973gs}, and stated more formally by Israel \cite{Israel:1986gqz}. Recently, Kehle and Unger \cite{Kehle:2022uvc} have proved that this conjecture is false for certain types of matter. They showed that it is possible for a massless charged scalar field to undergo spherically-symmetric gravitational collapse to form a black hole which is exactly Schwarzschild for a finite period of time but which then absorbs charge to become an exactly extremal Reissner-Nordstr\"om black hole after a finite time. They have constructed similar solutions for massless charged Vlasov matter \cite{Kehle:2024vyt}. They have also conjectured that it should be possible to form an extremal Kerr black hole in finite time starting from vacuum initial data. 

The proof of \cite{Kehle:2022uvc} is based on {\it characteristic gluing}, a method of constructing solutions to Einstein's equations first developed by Aretakis, Czimek and Rodnianski \cite{Aretakis:2021fel,Aretakis:2021fgz,Aretakis:2021lzi,Czimek:2022max}. Two explicitly-known spacetimes are glued along a common characteristic (i.e.~null) hypersurface by choosing appropriate data on this hypersurface. At early/late time the data on this hypersurface coincides with that of the original spacetimes to be glued, with a non-trivial ``gluing'' region in between. The data in this region must be chosen such that the equations of motion hold on the surface and the metric and matter fields have the desired degree of smoothness, which requires that suitably many transverse derivatives of the equations of motion must also be satisfied on the surface. In general, this procedure requires fine-tuning of the data in the gluing region. 

In the construction of \cite{Kehle:2022uvc}, the characteristic gluing hypersurface is the event horizon. In the simplest case, it is constructed by gluing a null cone in Minkowski spacetime to the event horizon of an extremal Reissner-Nordstr\"om black hole (a more complicated 2-stage gluing is required to ensure an intermediate Schwarzschild phase). In spherical symmetry, the free data is simply the scalar field profile in the gluing region, i.e.~the scalar field profile on a section of the event horizon. Kehle and Unger use an Ansatz for this profile consisting of a superposition of a finite number of disjoint bump functions, with each bump multiplied by an arbitrary parameter. They proved that there exists a choice for these parameters that achieves gluing, provided that the mass $M$ of the final black hole is large in units of the scalar field charge $\e$, i.e.~$\e M \gg 1$. Once the characteristic data on the horizon has been constructed, general theorems can be used to prove that it arises from a spacetime describing gravitational collapse to produce a black hole that is exactly extremal Reissner-Nordstr\"om after a finite time.

Although very elegant, the nature of the proof does not reveal much about the actual form of the scalar field profile on the horizon. It is also unclear just how large $\e M$ has to be in order for the gluing construction to work. In this paper we will answer these questions by performing the gluing numerically, using several different Ans\"atze for the scalar field profile on the horizon. Some of these are of the form discussed by Kehle and Unger, i.e.~disjoint bumps, but others are not. For each Ansatz we determine the minimum value of $\e M$ required to achieve gluing. We find that this differs significantly for the different Ans\"atze. This situation is reminiscent of critical collapse \cite{Choptuik1993,Gundlach1997,Gundlach2007}, where the threshold amplitude depends sensitively on the initial profile while the critical solution itself—analogous here to the extremal black hole—remains universal.

A second question that we will investigate is how the inclusion of a non-zero mass for the scalar field affects the results. The examples of Kehle and Unger involve matter with vanishing mass-to-charge ratio. Kehle and Unger argued that their results will also hold for a charged scalar field or charged Vlasov matter with a non-zero but small mass-to-charge ratio. However, if the mass-to-charge ratio is not small enough then it has been proved that examples like these cannot exist. For matter satisfying a ``local mass-charge inequality'', asserting roughly that the mass-to-charge ratio of matter is greater than (or equal to) $1$ in suitable units, then it is not possible for an extremal Reissner-Nordstr\"om black hole to form in finite time in gravitational collapse \cite{Reall:2024njy} or from a pre-existing non-extremal black hole \cite{McSharry:2025iuz}.\footnote{Similar results hold for {\it supersymmetric} asymptotically anti-de Sitter black holes \cite{McSharry:2025iuz}.} It is not known whether this result is sharp, i.e.~if matter has mass-to-charge ratio strictly less than $1$ then can one construct a solution describing gravitational collapse to form an extremal Reissner-Nordstr\"om black hole in finite time? We will explore this question numerically for the case of a scalar field of mass $\m$ and charge $\e$. For each of our scalar field Ans\"atze, we will show that if $\m/\e$ is not too large (well below $1$ but beyond the regime of validity of the methods of Kehle and Unger) then the gluing can be performed and so there exists a solution describing gravitational collapse to form an extremal Reissner-Nordstr\"om black hole in finite time. 

A final topic that we will investigate is the generalisation of the gluing construction to allow for a non-vanishing cosmological constant $\Lambda$, either positive or negative. We will show that the gluing construction works for any value of the cosmological constant. We find that when $\Lambda$ is negative, the minimum value of $\e M$ needed to reach extremality is larger than for $\Lambda=0$, whereas when the cosmological constant is positive $\e M$ can be smaller. 

An important feature of the characteristic gluing process is that it produces spactimes with a finite degree of differentiability. The proof of Kehle and Unger establishes that one can construct third-law violating spacetimes of any desired degree of differentiability. To achieve gluing in $C^k$ requires solving for $2k+1$ real parameters. This makes finding numerical solutions increasingly difficult as $k$ increases, since it requires scanning over a higher dimensional space of parameters. Therefore we will limit ourselves to gluing constructions of fairly low differentiability ($C^0$, $C^1$ and $C^2$). However, we do not expect that increasing the differentiability further would lead to qualitatively new insights. Another feature of the gluing construction is that it makes no qualitative difference whether or not the final black hole is extremal: from this perspective an extremal black hole is no more fine-tuned than a black hole with any prescribed charge-to-mass ratio. So in much of this paper we will also discuss gluing to non-extremal black holes.

This paper is structured as follows. In Section \ref{sec:gluing_theory}, we describe characteristic gluing, introduce the Einstein-Maxwell-charged scalar field system, and provide the corresponding governing equations in spherical symmetry. We discuss gauge freedom and how it can be exploited in our search for gluing solutions. We begin Section \ref{sec:results} by briefly reviewing the results of Kehle and Unger \cite{Kehle:2022uvc}. We then describe our numerical results, starting with the lowest regularity class, $C^0$. We later explore $C^1$ gluing, discuss the $C^2$ case, and compare our findings of each case. In Section \ref{sec:discussion}, we summarise and discuss our results and some possible future directions.

\section{Characteristic Gluing to Reissner-Nordstr\"om Spacetimes}\label{sec:gluing_theory}

\subsection{Introduction to gluing}

\label{subsec:gluing_intro}

Suppose we would like to construct a spacetime that contains a region that is (conformally) flat, and a region that is the exterior region of a stationary black hole. One approach is to use characteristic gluing: assume the two regions lie on a common null cone, and choose the metric functions and the matter fields to interpolate appropriately between the two regions along this cone. Then, by exploiting the well-posedness of the characteristic initial value problem \cite{Luk:2011vf}, one can extend the solution to a slab of finite thickness around the cone, thereby obtaining a spacetime with the desired properties.

\begin{figure}[t]
    \centering
    \includegraphics[width=0.6\linewidth]{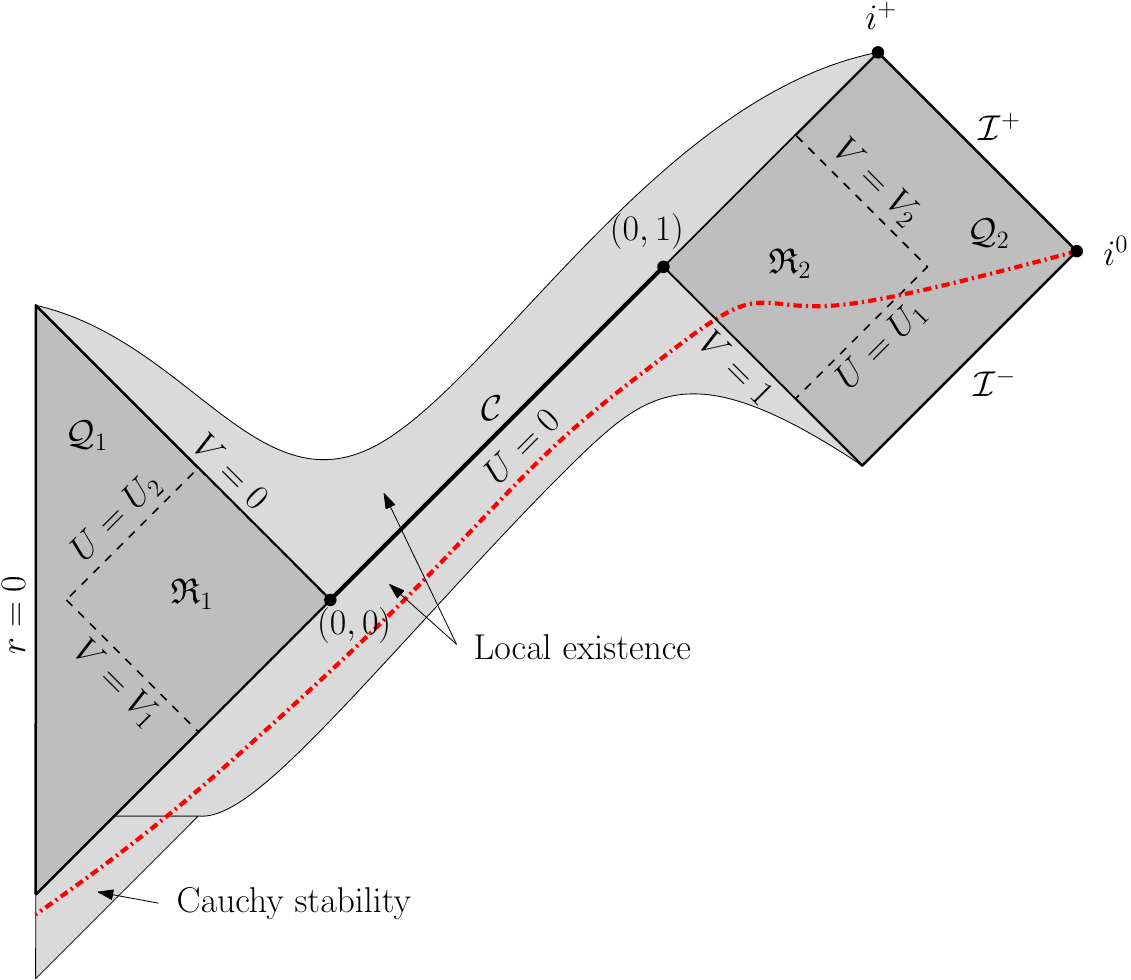}
    \caption{Setup for characteristic gluing with zero cosmological constant along the outgoing null cone $\mathcal{C}\equiv\{0\}\times[0,1]$. The metric and matter content in regions $\mathfrak{R}_1$ and $\mathfrak{R}_2$ are known and are given by $\mathcal{Q}_1$ and $\mathcal{Q}_2$, respectively. Data must be imposed on $\mathcal{C}$ such that the Einstein equations are satisfied. In the spherically-symmetric situation considered in this paper, $\mathfrak{R}_1$ can be extended to $r=0$, the centre of symmetry, to obtain the dark triangular region. The rectangle $\mathfrak{R}_2$ can be extended to get a complete future null infinity $\mathcal{I}^+$, as well as a segment of past null infinity $\mathcal{I}^-$, and spacelike ($i^0$) and future timelike ($i^+$) infinity. Once a valid solution to the characteristic gluing problem is found, the well-posedness of the characteristic initial value problem \cite{Luk:2011vf} can be exploited to obtain a local extension of spacetime. To extend this local solution to $r=0$, one can use Cauchy stability \cite{Kehle:2022uvc}. A Cauchy surface is shown in red.}
    \label{fig:gluing_diagram}
\end{figure}

To make this idea more precise, let us introduce some notation.\footnote{We have borrowed some of the notation of \cite{Kehle:2022uvc}. See also their excellent introduction on characteristic gluing, and references therein.} Let $U$, $V$, be outgoing and ingoing null coordinates. Let $\mathfrak{R}_1=[0,U_2]\times[V_1,0]$ and $\mathfrak{R}_2=[U_1,0]\times[1,V_2]$ be rectangular regions, isometric to a subset of some spacetime $\mathcal{Q}_1$, $\mathcal{Q}_2$, respectively, for some $U_1<0<U_2$ and $V_1<0<1<V_2$ (see Figure \ref{fig:gluing_diagram}). The two regions, by construction, touch the outgoing null cone $\mathcal{C}\equiv \{(U,V)\, | \, U=0, V\in [0,1]\}$. In each of $\mathfrak{R}_1$, $\mathfrak{R}_2$, the spacetime is known exactly. Our goal now is to determine what values the metric and matter fields should take in the other regions---in particular, on $\mathcal{C}$.

The Einstein equations impose constraints in the form of evolution equations along $\mathcal{C}$. For instance, Raychaudhuri's equation will dictate how the expansion must evolve. Other equations will propagate other quantities. To obtain a successful gluing procedure in the $C^k$ regularity class, we must ensure that the metric, the matter fields and all their derivatives to $k$-th order are continuous all along $\mathcal{C}$. The initial conditions at $(U=0,V=0)$, including the transverse derivatives $\partial_U(\cdot)$, are known, as these can be computed from $\mathcal{Q}_1$ in $\mathfrak{R}_1$. Similarly, the final values of all of these quantities at $(0,1)$ are known from $\mathcal{Q}_2$. Thus, the problem has now become a shooting problem: what profiles for the metric and matter fields must be imposed on $\mathcal{C}$ so that the correct values are obtained at the sphere $(U=0,V=1)$ of $\mathfrak{R}_2$ after integration of the evolution equations?

In the problem we are considering in this paper, $\mathcal{Q}_1$ is a conformally flat spacetime, i.e.~Minkowski, de Sitter, or anti-de Sitter, depending on the sign of the cosmological constant. In the case of de Sitter, we assume that $\mathcal{Q}_1$ is isometric to a subset of the static patch of de Sitter spacetime. Thus, for any sign of the cosmological constant, $\partial_Ur<0$ in $\mathfrak{R}_1$, where $r=r(U,V)$ is the area-radius. The rectangle  $\mathfrak{R}_1$ can be extended up to the centre of symmetry, $r=0$, thereby obtaining the dark triangular region on the left of Figure \ref{fig:gluing_diagram}. The spacetime inside this triangle is isometric to (a subset of) $\mathcal{Q}_1$ with the property that $\partial_U r<0$, i.e.~there are no antitrapped surfaces. 

Meanwhile, $\mathcal{Q}_2$ is a (possibly extremal) black hole spacetime in the Reissner-Nordstr\"om class, with asymptotics depending on the sign of the cosmological constant. One edge of $\mathfrak{R}_2$ is isometric to the event horizon; the rest is isometric to the exterior region. One can extend $\mathfrak{R}_2$ to the ``north east'' and ``south east'' (refer to Figure \ref{fig:gluing_diagram}). When the cosmological constant vanishes, $\mathcal{Q}_2$ is asymptotically flat, and the extension of $\mathfrak{R}_2$ inherits this property. In particular, the full, dynamical spacetime gains a complete future null infinity $\mathcal{{I}^+}$, as well as future timelike infinity $i^+$ and spacelike infinity $i^0$. It also obtains a past null infinity $\mathcal{I}^-$, although only part of it is isometric to that of $\mathcal{Q}_2$; the rest depends on the data on $\mathcal{C}$. If instead the cosmological constant is negative, then $\mathfrak{R}_2$ can be extended up to the timelike boundary of anti-de Sitter spacetime, whereas if it is positive, null infinity will be replaced by a cosmological horizon.

Once $\mathfrak{R}_1$ and $\mathfrak{R}_2$ have been appropriately glued, the null cone $\mathcal{C}$ can be extended into a full $4$-dimensional region of spacetime, thanks to a result of Luk \cite{Luk:2011vf}. Taking first the two null hypersurfaces $\{(U,V)\, | \, U\geq0, V=0\}$ and $\{(U,V)\, | \, U=0, V\geq0\}$, one can locally extend the spacetime to the north. Next, taking the hypersurfaces $\{(U,V)\, | \, U\leq0, V=1\}$ and $\{(U,V)\, | \, U=0, V\leq1\}$, one can locally extend the spacetime to the south. Extending this segment to the centre, $r=0$, is non-trivial, but can be achieved by exploiting Cauchy stability \cite{Kehle:2022uvc}. Note that, in spherical symmetry, the extended spacetime solution loses a derivative at the centre compared with the characteristic data \cite{Kehle:2022uvc}. Thus, if we wish to obtain a spacetime of regularity $C^k$, we require $C^{k+1}$ characteristic data.

With the spacetime finally constructed, one can then choose a spacelike Cauchy surface starting from $i^0$, crossing into the local existence region to the past of $\mathcal{C}$, and terminating at the centre of symmetry to the past of the region isometric to $\mathcal{Q}_1$, as shown in Figure \ref{fig:gluing_diagram}. Taking this surface and the induced metric, extrinsic curvature, and matter fields on it, and now abstracting away the rest of the spacetime, one has an initial value problem which does not contain a black hole initially, but whose maximal global hyperbolic development is a black hole solution. In particular, at late time the exterior region and the horizon are \emph{exactly} Reissner-Nordstr\"om. Thus this solution describes gravitational collapse to form an exactly Reissner-Nordstr\"om black hole in finite time. 

\begin{figure}
    \centering
    \includegraphics[width=0.75\linewidth]{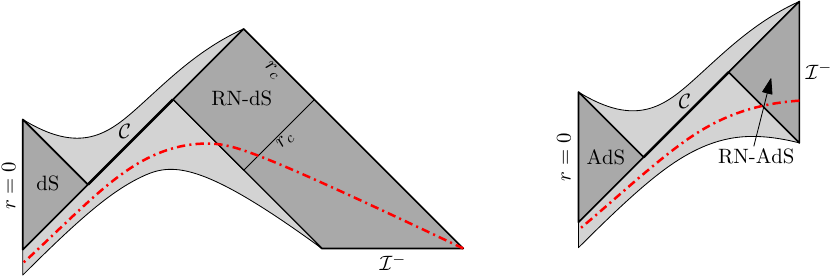}
    \caption{\textit{Left:} Characteristic gluing with a positive cosmological constant. The Cauchy slice (red) is complete and extends to past null infinity. \textit{Right:} Characteristic gluing with a negative cosmological constant. The Cauchy slice extends to the timelike boundary.} 
    \label{fig:gluing_diagram_cosmo}
\end{figure}

When the cosmological constant is positive, $\mathcal{I}^\pm$ in Figure \ref{fig:gluing_diagram} are replaced by the cosmological horizons of the Reissner-Nordstr\"om-de Sitter spacetime. In this case we can take an incomplete spacelike slice from the cosmological horizon, down the local existence region to the past of the black hole horizon, and down to $r=0$. Alternatively, the slice could start at past infinity in the Reissner-Nordstr\"om-de Sitter region instead of at the cosmological horizon. This scenario is shown on the left of Figure \ref{fig:gluing_diagram_cosmo}. In the case of a negative cosmological constant, one can then take a spacelike slice from $r=0$ extending out to timelike infinity in the Reissner-Nordstr\"om-anti-de Sitter region, as illustrated on the right of Figure \ref{fig:gluing_diagram_cosmo}. The solution constructed this way is independent of the boundary conditions at timelike infinity (unless one wishes to evolve it to earlier times). 

In this paper, we will focus on determining the appropriate data to impose on $\mathcal{C}$, but will not discuss the extension to a full spacetime in any further detail, as this is already well explained in Section 5 of \cite{Kehle:2022uvc}.

\subsection{Einstein-Maxwell-Charged Scalar Field System}

Let us now make the discussion more concrete by introducing the system we will study: the Einstein-Maxwell-charged scalar field system. The equations of motion read \cite{Kehle:2022uvc,Rossetti:2023mrx}
\begin{subequations}\label{eq:EMCSF}
\begin{align}
    R_{ab}-\frac{1}{2}Rg_{ab}+\Lambda g_{ab}&=2\left(T^{\rm EM}_{ab}\, + \,T^{\rm CSF}_{ab}\right),\\
    \nabla^aF_{ab}&=2\,\e\,{\rm Im}(\Phi \overline{D_b\Phi}),\\
    g^{ab}D_aD_b\Phi&=\m^2\Phi,
\end{align}
\end{subequations}
where $\Lambda$ is the cosmological constant, $F={\rm d}\mathcal{A}$ is the electromagnetic tensor, $\e$ is the electromagnetic coupling constant (the charge of the complex scalar field $\Phi$), $D_a=\nabla_a+i \e \mathcal{A}_a$ is the gauge covariant derivative, and $\m$ is the mass parameter of the scalar field. The energy-momentum tensors are
\begin{subequations}
\begin{align}
    T_{ab}^{\rm EM} &=g^{cd}F_{cb}F_{da}-\frac{1}{4}F_{cd}F^{cd}g_{ab},\\
    T_{ab}^{\rm CSF} &={\rm Re}\left(D_a\Phi\overline{D_b\Phi}\right)-\frac{1}{2}g_{ab}\left(D_c\Phi \overline{D^c\Phi}+\m^2|\Phi|^2\right).
\end{align}
\end{subequations}
We will restrict to the spherically-symmetric case, and write the metric in the form
\begin{equation}\label{eq:spherical_metric}
    {\rm d}s^2=-\Omega^2{\rm d}U {\rm d}V+r^2{\rm d}\omega^2,
\end{equation}
where $\Omega$ and $r$ are functions of the two null coordinates $U$ and $V$ only, and $d\omega^2$ is the line element on the 2-sphere. We assume that $\Phi$, $\mathcal{A}$ and $F$ are also spherically symmetric, such that we have
\begin{equation}
    F = \frac{Q\,\Omega^2}{2r^2} {\rm d}U \wedge {\rm d}V,
\end{equation}
where $Q$ is another function of $U$ and $V$, the charge enclosed by the $2$-sphere with those values of $U,V$. We will fix the gauge to $\mathcal{A}=\mathcal{A}_UdU$. With these choices, \eqref{eq:EMCSF} becomes the following set of equations.

\begin{subequations}\label{eq:EMCSF_spherical}
\paragraph{Raychaudhuri's equations}
\begin{align}
    \partial_V\left(\frac{\partial_Vr}{\Omega^2}\right) &=-\frac{r}{\Omega^2}\left|\partial_V\Phi\right|^2 \label{eq:EMCSF_RayV} \\
    \partial_U\left(\frac{\partial_Ur}{\Omega^2}\right) &=-\frac{r}{\Omega^2}\left|D_U\Phi\right|^2 \label{eq:EMCSF_RayU}
\end{align}

\paragraph{Maxwell's equations}
\begin{align}
    \partial_VQ &= \e r^2\,{\rm Im}\left(\Phi \overline{\partial_V\Phi}\right) \label{eq:EMCSF_MaxwellV} \\
    \partial_UQ &= -\e r^2\,{\rm Im}\left(\Phi \overline{D_U\Phi}\right)\label{eq:EMCSF_MaxwellU} \\
    \partial_V\mathcal{A}_U &= -\frac{Q\Omega^2}{2r^2} \label{eq:EMCSF_MaxwellA}
\end{align}

\paragraph{Wave equations}
\begin{align}
    \partial_V\partial_Ur &= -\frac{\Omega^2}{4r}-\frac{\partial_Ur\,\partial_Vr}{r}+\frac{\Omega^2Q^2}{4r^3}+\frac{\m^2\Omega^2r\,|\Phi|^2}{4}+\frac{\Omega^2r\,\Lambda}{4} \label{eq:EMCSF_waveR}\\
    \partial_V\partial_U\log(\Omega^2) &=\frac{\Omega^2}{2r^2}+\frac{2\partial_Ur\,\partial_Vr}{r^2}-\frac{\Omega^2Q^2}{r^4}-2\,{\rm Re}\left(D_U\Phi\overline{\partial_V\Phi}\right)\\
    \partial_V\partial_U\Phi &=-\frac{\partial_U\Phi\,\partial_Vr+\partial_Ur\,\partial_V\Phi}{r}+\frac{i\e\Omega^2Q\,\Phi}{4r^2}-i\e\frac{\mathcal{A}_U\,\Phi\,\partial_Vr}{r}-i\e\mathcal{A}_U\,\partial_V\Phi-\frac{\Omega^2\m^2\, \Phi}{4} \label{eq:EMCSF_wavePhi}
\end{align}
\end{subequations}
Consider a solution of these equations for which the metric, Maxwell potential, and scalar field are $C^k$ functions of $U,V$. Hence $r$, $\Omega$, $\mathcal{A}_U$ and $\Phi$ are $C^k$ functions. Equations \eqref{eq:EMCSF_MaxwellV} and \eqref{eq:EMCSF_MaxwellU} then imply that $Q$ is also a $C^k$ function and equations \eqref{eq:EMCSF_RayV}, \eqref{eq:EMCSF_RayU} and \eqref{eq:EMCSF_waveR} imply that $r$ is actually a $C^{k+1}$ function, i.e.~smoother than initially assumed. Similarly, equation \eqref{eq:EMCSF_MaxwellA} implies that $\partial_V \mathcal{A}_U$ is a $C^k$ function. 

To obtain a $C^k$ solution, the quantities that need to be glued on $\mathcal{C}$ are $r$, $\Omega$, $Q$, $\mathcal{A}_U$ and $\Phi$, and all their partial derivatives to the order just described. As we will explain in Section \ref{sec:gauge_RN}, $\Omega$ can be set to $1$ along $\mathcal{C}$ without loss of generality by exploiting the freedom in defining $V$. Thus, after making an Ansatz for $\Phi(V)$ on $\mathcal{C}$, we can solve \eqref{eq:EMCSF_RayV} to find $r(V)$, with initial conditions supplied at either $V=0$ or $V=1$, depending on whether we wish to integrate forwards or backwards in $V$. We can now sequentially obtain $Q(V)$ from \eqref{eq:EMCSF_MaxwellV}, $\mathcal{A}_U(V)$ from \eqref{eq:EMCSF_MaxwellA} and $(\partial_Ur)(V)$, $(\partial_U\Omega)(V)$ and $(\partial_U\Phi)(V)$ from the wave equations. To obtain higher $U$ derivatives, we iteratively differentiate each equation with respect to $U$, substitute $\partial_UQ$ with the right-hand side of \eqref{eq:EMCSF_MaxwellU} if necessary, and solve sequentially as for lower orders.

\subsection{Gauge Freedom}\label{sec:gauge_RN}


There is some residual gauge freedom in our Ansatz \eqref{eq:spherical_metric} which we can use to simplify the problem \cite{Kehle:2022uvc}.
Firstly, we can rescale $U\to f(U)$, $V\to g(V)$, so that the $dU dV$ term in the metric becomes
\begin{equation}
    -\Omega^2(U,V) \to -\tilde{\Omega}^2(U,V)=-\Omega^2(f(U),g(V))\,f'(U)g'(V).
\end{equation}
We require $\tilde{\Omega}^2>0$, which implies that $f'(U)$ and $g'(V)$ have the same sign. In order to keep $\partial_U$ and $\partial_V$ pointing in the same direction as before we will choose the sign to be positive. We now use the freedom in $g(V)$ to set $\tilde{\Omega}=1$ along the gluing hypersurface $\mathcal{C}$. Henceforth we will assume $g(V)$ is fixed by this condition.

Now consider the transverse derivatives of $\Omega$ and $r$ under the above gauge transformation, evaluated at the sphere $(U,V)=(U_0,V_0)$,
\begin{align}
     \partial_U(\Omega^2)(U_0,V_0) &\to \partial_{f(U)}(\tilde{\Omega}^2)(U_0,V_0) &&\propto [2\Omega(U_0,V_0)\,\partial_U\Omega(f(U_0),g(V_0))]\,f'(U_0)g'(V_0) \nonumber \\& &&\qquad\qquad+\Omega^2(f(U_0),g(V_0))\,f''(U_0)g'(V_0) \\
    \partial_Ur(U_0,V_0) &\to \partial_{f(U)}\tilde{r}(U_0,V_0) &&\propto \partial_Ur(f(U_0),g(V_0))\,f'(U_0).
\end{align}
We have two free parameters, $f'(U_0)$ and $f''(U_0)$, which can be used to simultaneously rescale $\partial_{f(U)}\tilde{r}(U_0,V_0)$ by any positive number (recall $f'(U_0)>0$) and bring $\partial_{f(U)}(\tilde{\Omega}^2)(U_0,V_0)$ to any desired value. In particular, we can set $\partial_{f(U)}\tilde{\Omega}(U_0,V_0)$ to zero.

By choosing the higher derivatives of $f$ appropriately, we can specify all transverse derivatives $\partial_U^n\Omega$ to any given order, e.g.~we can choose a gauge where $\Omega=1$ and all its derivatives (in both $U$ and $V$) vanish to order $k$ on the sphere $(U_0,V_0)$. The second Raychaudhuri equation, \eqref{eq:EMCSF_RayU}, guarantees that the second and higher order transverse derivatives of $r$ are all consistent across the different gauge choices of $\Omega$.

There is also a residual electromagnetic gauge freedom $\mathcal{A}_U\to \mathcal{A}_U+2\partial_U\chi$ along with $\Phi\to\Phi e^{4i\chi}$, where $\chi$ is only a function of $U$ such that we remain in the gauge $\mathcal{A}_V=0$. Including the gauge freedom from the rescaling of the null coordinates, we have
\begin{equation}
    \mathcal{A}_U(U_0,V_0)\to\tilde{\mathcal{A}}_U(U_0,V_0)=f'(U_0)\,\mathcal{A}_U(f(U_0),g(V_0))+2f'(U_0)\,\partial_U\chi (f(U_0),g(V_0)).
\end{equation}
Thus, we can set $\tilde{\mathcal{A}}_U(U_0,V_0)=0$ by an appropriate choice of $\chi$. Similarly, the transverse derivatives can be made to vanish by choosing the higher derivatives of $\chi$ carefully.

It will be useful now to introduce the Hawking mass,
\begin{equation}
    m=\frac{r}{2}\left(1+\frac{4\partial_Ur\partial_Vr}{\Omega^2}-\frac{\Lambda\, r^2}{3}\right).
\end{equation}
From the above, and using \eqref{eq:EMCSF_MaxwellA}, it is easy to see that
\begin{align}
    m &\to \frac{r}{2}\left[1+\frac{f'(U_0)\partial_Ur\, g'(V_0)\partial_Vr}{f'(U_0)g'(V_0)\Omega^2}-\frac{\Lambda\, r^2}{3}\right]=m\\
    Q &\to -2r^2\frac{g'(V_0)\partial_V{\mathcal{A}_U}f'(U_0)}{f'(U_0)g'(V_0)\Omega^2} = Q
\end{align}
so the Hawking mass and the charge are gauge-invariant, as expected.

The characteristic gluing procedure glues two spheres that lie on a common null hypersurface $\mathcal{C}$. In other words, it interpolates between the data on each sphere. Due to the spherical symmetry of the spacetime, the data on a sphere of constant $U$, $V$ is constant all over the sphere. Thus, each quantity to be glued (e.g.~$\Phi$) takes a single value on each such sphere. We will call the set of values of $r$, $\Omega$, $m$, $Q$, $\mathcal{A}_U$, $\Phi$, and their derivatives on a given sphere a \emph{sphere data set}. When performing characteristic gluing, say between a Minkowski region and a Reissner-Nordstr\"om black hole, we need only glue data that are gauge-equivalent to Minkowski and Reissner-Nordstr\"om data, respectively. To understand what sphere data sets are gauge-equivalent to each other, it is easiest to start from a so-called ``lapse-normalised'' gauge (i.e.~with $\Omega\equiv 1$) and then perform a gauge transformation. 

\paragraph{Minkowski}

We can write Minkowski spacetime in the lapse-normalised gauge as
\begin{equation}
    {\rm d}s^2=-{\rm d}U {\rm d}V+r^2{\rm d}\omega^2,
\end{equation}
where $r(U,V)=(V-U)/2\in[0,\infty)$. Thus we can read off $\Omega\equiv1$, $\partial_Ur=-1/2$, $\partial_Vr=1/2$, $\mathcal{A}_U=0$, $\Phi\equiv0$, and of course the Hawking mass and the charge both vanish. After performing a gauge transformation of the type described above, we can bring $\partial_Ur$ to any negative value. Since $m$ is gauge-invariant, $\partial_Vr$ must transform such that $m$ remains zero. Similarly, we can transform $\mathcal{A}_U$ to any desired value, but $Q$ must remain zero. We therefore obtain the following result \cite{Kehle:2022uvc}:

\begin{prop}
    A sphere data set with $r=R>0$, $\partial_Ur<0$, $m=0$, $Q=0$, $\partial_U^n\Phi=0$ and $\partial_V^n\Phi=0$ for $n=0,\dots,k$ is gauge-equivalent to a lapse-normalised data set of a Minkowski sphere.
\end{prop}

Notice that there is no gauge-independent condition on $\Omega$. A similar result holds for spheres in de Sitter (dS) and anti-de Sitter (AdS) spacetimes. In all cases we have $\partial_Ur<0$ and $m=0$; for de Sitter we will assume the sphere lies within the cosmological horizon, i.e.~$1-\Lambda r^2/3>0$ (this is to ensure $\partial_Vr>0$, i.e.~that the sphere is not trapped). The above sphere data sets will be used to set the values of the metric functions and scalar field at $V=0$.

\paragraph{Reissner-Nordstr\"om Horizon}

We can repeat the same exercise for the Reissner-Nordstr\"om spacetime. We denote the mass of the black hole by $M$ and its dimensionless charge by 
\begin{equation}
    \q = \frac{Q}{Q_{\rm max}(M,\Lambda)}
\end{equation}
where $Q_{\rm max}$ is the largest possible charge of the black hole. For $\Lambda =0$ we have $Q_{\rm max} = M$ and the horizon radius is $r_+=M(1+\sqrt{1-\q^2})$. If $\Lambda\neq0$ then $Q_{\rm max}$ is determined by solving the simultaneous equations $f(r_+)=0$, $f'(r_+)=0$, where $f(r)=1-2M/r+Q^2/r^2-\Lambda r^2/3$. 

A lapse-normalised sphere on the horizon of a Reissner-Nordstr\"om black hole has $\Omega\equiv 1$, $r=r_+$, $\partial_U r=-1/2$, $\partial_V r=0$, $Q=\q M$, $\Phi\equiv0$. Hence we have \cite{Kehle:2022uvc}:

\begin{prop}
    A sphere data set with $r=r_+$, $\partial_Ur<0$, $\partial_Vr=0$, $Q=\q M$, $\partial_U^n\Phi=0$ and $\partial_V^n\Phi=0$ for $n=0,\dots,k$ is gauge-equivalent to a lapse-normalised data set on the horizon sphere (the sphere with $r=r_+$ and constant $U$ and $V$) of a Reissner-Nordstr\"om spacetime. 
\end{prop}

This is essentially Birkhoff's theorem. Analogous results hold for the Reissner-Nordstr\"om-(A)dS spacetimes. The condition $\partial_Ur<0$ ensures that the Reissner-Nordstr\"om horizon sphere in question is on the black hole horizon rather than the white hole horizon.

\subsection{Scalar field Ans\"atze}\label{sec:profiles}

We wish to find a profile for $\Phi$ along $\mathcal{C}$ that interpolates between a Minkowski sphere at $V=0$ and a Reissner-Nordstr\"om horizon sphere at $V=1$. Thus $\Phi$ must be chosen such that after solving the evolution equations \eqref{eq:EMCSF_spherical} we achieve $r(V=1)=r_+$, $\partial_Vr(1)=0$, $Q(1)=Q$, and $\partial_U^n\Phi(1)=0$, $n=1,\dots,k$, starting with the Minkowski/dS/AdS values at $V=0$. We also require that $\inf r>0$ and $\sup \partial_Ur<0$ for $V \in [0,1]$. The former condition is required for a well-defined metric, the latter ensures that no antitrapped sphere is present on the gluing surface (to avoid the possibility of obtaining a solution where the gluing occurs inside a white hole). We set $\Omega=1$ on $\mathcal{C}$ using the gauge freedom described above. We do not need to glue $\partial_U^2r$, $\partial_U\Omega$, $\partial_UQ$, $\partial_U\mathcal{A}_U$, and their transverse derivatives, as these can all be gauged away (by the Propositions above). Note that we must still evolve these quantities as they appear in the equations controlling quantities that must be glued.

In practice, as in \cite{Kehle:2022uvc}, we begin by first solving the Raychaudhuri equation \eqref{eq:EMCSF_RayV} {\it backwards} from $V=1$ to $V=0$ (using the initial conditions $r(1)=r_+$, $\partial_Vr(1)=0$). Then, we evolve the remaining equations, \eqref{eq:EMCSF_MaxwellV}--\eqref{eq:EMCSF_wavePhi}, forwards in time. The initial condition for $\partial_Ur(0)$ can be derived from the requirement that the Hawking mass be zero, giving 
\begin{equation}
\label{eq:dUr(0)}
    \partial_Ur(0) = \frac{\Lambda\, r(0)^2/3-1}{4\partial_Vr(0)}.
\end{equation}
The initial conditions for the remaining quantities ($\partial_U\Omega$, $\partial_U^2r$, $\partial_U Q$, etc.) can all be set to zero using the gauge freedom described above. At the end we extract the final values of $\partial_U^n\Phi$ and $Q$, and also check whether $\inf r>0$ and $\sup \partial_Ur<0$. 

Without loss of generality, let us assume that $\e M>0$. Suppose that we wish to obtain a black hole with $\q>0$. Following \cite{Kehle:2022uvc}, we choose an Ansatz for $\Phi$ of the form
\begin{equation}
    \Phi(V) = \rho(V)e^{-iV},
\end{equation}
where $\rho$ is a compactly-supported smooth real-valued function in the interval $[0,1]$. This choice for the phase ensures that $\partial_VQ>0$, and hence that we will reach a positive charge, as desired. If instead we wished to reach $\q<0$, we would swap the sign of the phase, i.e.~$\Phi(V)=\rho(V)e^{iV}$. For $\q=0$, we would remove the phase altogether, so that $\Phi$ becomes purely real.

There is of course an enormous amount of freedom available in choosing $\rho$. We will consider several different Ans\"atze of the form
\begin{equation}\label{eq:rho_def}
    \rho(V) = \sum_{j=1}^{2k+1}\alpha_j\chi_j(V),
\end{equation}
where $\chi_j$ are smooth functions vanishing at $V=0,1$ and $\alpha_j$ are real coefficients. We have $2k+1$ parameters as this is the number of constraints that we must shoot for: one for the final charge of the black hole, and $2k$ for the real and imaginary parts of the first $k$ transverse derivatives of $\Phi$. We will refer to the set of functions $\chi_j$ as a basis. 

We have mainly experimented with three different choices for the basis functions $\chi_j$. The first two, inspired by \cite{Kehle:2022uvc}, consist of compactly-supported bump functions with disjoint support. More specifically, in the region they are non-zero, these functions are
\begin{align}
    \chi^e_j(V) &= N_e(K)\,e^\frac{1}{[V-(j-1)/K][V-j/K]},\quad \frac{j-1}{K}<V<\frac{j}{K}\quad && \text{(even bump)},\label{eq:even_Ansatz} \\
    \chi^o_j(V) &= N_o(K)\,\left[1-\frac{K^2(1-2j+2KV)^2}{2(j-KV)^2(1-j+KV)^2}\right]e^\frac{1}{[V-(j-1)/K][V-j/K]},\quad \frac{j-1}{K}<V<\frac{j}{K}\quad && \text{(odd bump)},\label{eq:odd_Ansatz}
\end{align}
where $K=2k+1$ is the total number of parameters, $j=1,\dots,K$, and $N_e(K)$, $N_o(K)$ are $K$-dependent constants that normalise the height of each bump to $1$. Note that the ``odd Ansatz'', $\chi^o_j$, is simply the derivative of the ``even Ansatz'', $\chi^e_j$. The third basis consists of the set of polynomials 
\begin{equation}
    V^{K+1}(1-V)^{K+1}V^j,\qquad j=0,\dots,K-1,
\end{equation}
orthonormalised by the Gram-Schmidt procedure. We will call it the ``polynomial Ansatz''. Notice that in each case, the Ansatz for $\rho(V)$ is such that all its derivatives to $k$-th order vanish at both endpoints $V\in\{0,1\}$.

Note that the final mass of the black hole, $M$, introduces a length scale, so we can form the following dimensionless parameters: $\q$, $\e M$, $\m/\e$, and $\Lambda M^2$. We will be using these henceforth.
Having now chosen a profile for the scalar field, we can provide an initial guess for $\alpha_1,\dots,\alpha_{2k+1}$, integrate the differential equations, and extract $Q(1)$ and the real and imaginary parts of $\partial_U^n\Phi(1)$, $n=1,\dots,k$. The aim now is to adjust the coefficients $\alpha_j$ (for fixed $\q$, $\e M$, $\m/\e$ and $\Lambda M^2$) to achieve $Q(1)=Q$ and $\partial_U^n\Phi(1)=0$. This can be done numerically; we employed the so-called ``good'' Broyden method for this purpose. Further details about our numerical methods are provided in Appendix \ref{sec:numerics}.

\section{Numerical results}\label{sec:results}

\subsection{General approach}\label{sec:gen_approach}

Following \cite{Kehle:2022uvc}, we will write the set of parameters $\alpha_j$ as a vector $\alpha=(\alpha_1,\dots,\alpha_{2k+1})\in\mathbb{R}^{2k+1}$ and denote its norm by $\beta\equiv|\alpha|$. We will also denote the unit-norm vector pointing in the same direction as $\alpha$ by $\hat{\alpha}\in S^{2k}$, so that we have $\alpha=\beta \hat{\alpha}$. $C^k$ gluing solutions are constructed as follows. 

First, the Raychaudhuri equation \eqref{eq:EMCSF_RayV} (with $\Omega=1$) is integrated backwards starting at $V=1$, with $r(1)=r_+$ and $\partial_V r(1)=0$, to determine $r(V)$ on $\mathcal{C}$. This equation is linear in $r$, so if we substitute $r = \xi r_+$ then the solution for $\xi$ depends only on the parameters $\alpha_i$ and not on $M,\q,\e,\m$ or $\Lambda$. The condition $r>0$ is equivalent to $\xi>0$ so we demand that the parameters $\alpha$ lie in the region of $\mathbb{R}^{2k+1}$ where this condition is satisfied. We will refer to this as the {\it restricted parameter space}. The results of Kehle and Unger establish that the restricted parameter space contains a round ball centered on the origin, corresponding to small $\beta$. We will be interested in exploring the restricted parameter space outside this ball.

Next, we integrate \eqref{eq:EMCSF_MaxwellV} forwards (with $Q(0)=0$) to determine $Q(V)$ on $\mathcal{C}$ and hence $Q(1)$. Demanding that this matches the black hole charge $Q$ gives
\begin{equation}
\label{Ieq}
   \frac{Q}{\e r_+^2} = I(\alpha) \quad \qquad I(\alpha) \equiv \int_0^1 {\rm d}V\,\xi^2 \,{\rm Im} (\Phi \overline{\partial_V \Phi} )\,.
\end{equation}
Note that $I$ depends only on the parameters $\alpha$ and not on $M,Q,\e,\m$ or $\Lambda$. We now fix $\hat{\alpha}$ and try to solve this equation for $\beta$ such that the resulting $\alpha$ belongs to the restricted parameter space. When the Ansatz is based on compactly supported $\chi_i$ and $\e r_+^2/Q$ is large then an argument of Kehle and Unger establishes that a (small) solution $\beta_Q(\hat{\alpha})$ exists for any $\hat{\alpha}$. The resulting set of points defines a surface of topology $S^{2k}$ in the restricted parameter space:
\begin{equation}\label{eq:Q_set}
    \mathfrak{Q}^{2k}\equiv\{\beta_Q(\hat{\alpha})\hat{\alpha}\,|\,\hat{\alpha}\in S^{2k}\}\subset \mathbb{R}^{2k+1}.
\end{equation}
When $\e r_+^2/Q$ is not large we will see below that a solution $\beta_Q$ may not exist for some or all values of $\hat{\alpha}$. Nevertheless, we can still define $\mathfrak{Q}^{2k}$ as above, with $\hat{\alpha}$ restricted to values for which solutions exist, but this surface might not be topologically $S^{2k}$, or even compact. More precisely $\mathfrak{Q}^{2k}$ is defined as the surface of constant $I$ in the restricted parameter space defined by \eqref{Ieq}. 

The next step is to determine $\partial_U r$ on $\mathcal C$ by integrating \eqref{eq:EMCSF_waveR} forwards, with $\partial_U r(0)$ determined by \eqref{eq:dUr(0)}. When $\Lambda=\m=0$ and $\e r_+^2/Q$ is large, Kehle and Unger proved that this procedure gives $\partial_Ur<0$ for every $\alpha\in\mathfrak{Q}^{2k}$. In Appendix \ref{sec:dur_cosmo_proof} we extend this result to the case of non-vanishing $\Lambda$. This ensures that there are no antitrapped surfaces on $\mathcal{C}$. This is necessary to exclude the possibility that the gluing occurs inside a white hole. In general we do not need $\partial_U r<0$ everywhere on $\mathfrak{Q}^{2k}$, but we will need to check that this condition is satisfied for the values of $\alpha$ we obtain from the next step. 

The final step is to solve the transport equations for $\partial_U^n \Phi(V)$, $n=1,2,\ldots,k$ along $\mathcal{C}$, starting from $\partial_U^n \Phi(0)=0$, and try to choose the parameters $\hat{\alpha}_i$ to satisfy $\partial_U^n \Phi(1)=0$, which is $2k$ real equations. For $\Lambda=\m=0$ and large $\e r_+^2/Q$, Kehle and Unger prove existence of a solution by exploiting their result that $\mathfrak{Q}^{2k}$ is topologically $S^{2k}$ along with the fact that the equations to be solved are odd under $\hat{\alpha} \rightarrow -\hat{\alpha}$ (which arises from the $\Phi \rightarrow -\Phi$ symmetry of the theory). The Borsuk-Ulam theorem then guarantees existence of (at least) two solutions related by the symmetry. More generally, $\mathfrak{Q}^{2k}$ is not topologically $S^{2k}$ so this argument does not apply and we will need to determine numerically whether or not a solution of these equations exists. If a solution does exist then we will refer to it as a {\it candidate solution}. The final step is to check that the {\it limiting condition} $\partial_U r<0$ is satisfied for this solution. If it is, then we will refer to the solution as a {\it valid solution}.

\subsection{$C^0$ Gluing}\label{sec:C0}

\subsubsection{Candidate solutions}

The difficulty of finding solutions to the equations of $C^k$ characteristic gluing increases with $k$. Therefore it makes sense to start with the simplest case, namely $C^0$ gluing. One might not expect this to be very physical, especially in view of the even lower regularity at the origin of the resulting solution (see Section \ref{subsec:gluing_intro}). However, as we will see, the $k=0$ case actually exhibits most of the important features that occur for higher $k$.

\begin{figure}
    \centering
    \includegraphics[width=\linewidth]{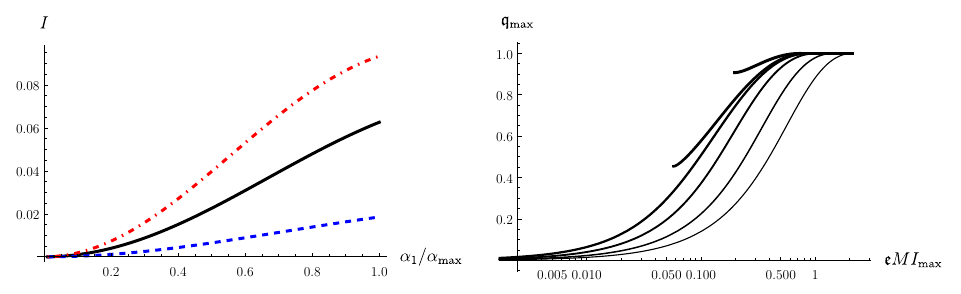}
    \caption{\textit{Left:} The function $I$ for the even (solid black), odd (dashed blue) and polynomial (dash-dotted red) Ans\"atze. \textit{Right:} $\q_{\rm max}$ against $\e M I_{\rm max}$ for different values of $\Lambda M^2$: from thickest to thinnest, $\Lambda M^2=0.17$, $0.12$, $0.1$, $0$, $-0.5$, $-2$. Notice that for $\Lambda M^2>1/9$, there is a lower bound on the charge of the final black hole.}
    \label{fig:RN_C0_I}
\end{figure}

In $C^0$ gluing, the Ansatz for the scalar field on $\mathcal{C}$ involves a single free parameter $\alpha_1$. The first step is to determine the restricted parameter space defined above. Numerically we find that this is given by $|\alpha_1| < \alpha_{\rm max}$ where $\alpha_{\rm max} = 0.5933$, $0.2987$ and $0.4343$ for the even, odd and polynomial Ans\"atze, respectively. For each $\alpha_1$ in the restricted parameter space we can calculate $I(\alpha)$; the result is given in the first panel of Figure \ref{fig:RN_C0_I}. We find that $I_{\rm max}\equiv \sup_\alpha I$ takes the values $I_{\rm max} = 0.0627$, $0.0186$ and $0.0934$ for the even, odd and polynomial Ans\"atze, respectively. In $C^0$ gluing we do not need to match derivatives of the scalar field and so the only equation to be solved by a candidate solution is  \eqref{Ieq}. Note that $\m$ does not enter this equation so $\m$ does not affect candidate solutions of $C^0$ gluing. This equation admits a solution if, and only if,
\begin{equation}
\label{eMbound}
    \frac{QM}{r_+^2} <\e M I_{\rm max}\,.
\end{equation}
For each of our Ans\"atze, $I(\alpha)$ is monotonic so there exists exactly one positive solution $\alpha_1$ if this condition is satisfied. The LHS is a known dimensionless function of the two dimensionless quantities $\q$ and $\Lambda M^2$. A plot of the maximum attainable (dimensionless) charge, $\q_{\rm max}$, against $\e M I_{\rm max}$ is shown in the second panel of Figure \ref{fig:RN_C0_I}.\footnote{In a Reissner-Nordstr\"om-de Sitter spacetime, the maximum value of the cosmological constant (determined by the condition that the black hole and cosmological horizons coincide) increases with the black hole charge from $\Lambda M^2=1/9$ at $\q=0$ to $\Lambda M^2=2/9$ at $\q=1$. This implies that for some fixed $\Lambda M^2>1/9$, the charge of the final black hole is bounded below. The effect of this is visible in the second panel of Figure \ref{fig:RN_C0_I}.} Hence for given $\e M$ we can easily determine whether or not a candidate solution exists for any value of these parameters. Turning this around, this equation determines the minimum value of $\e M$ required to construct a candidate solution with given $\q$ and $\Lambda M^2$. For example with $\q=1$ (extremality) and $\Lambda=0$ this gives $(\e M)_{\rm min} = 1/I_{\rm max}$, which is $15.96$, $53.86$, and $10.71$, for the even, odd, and polynomial Ans\"atze respectively (see also Table \ref{tab:results}).

\subsubsection{Massless scalar field}\label{sec:C0_massless}

\begin{figure}
    \centering
    \includegraphics[width=\linewidth]{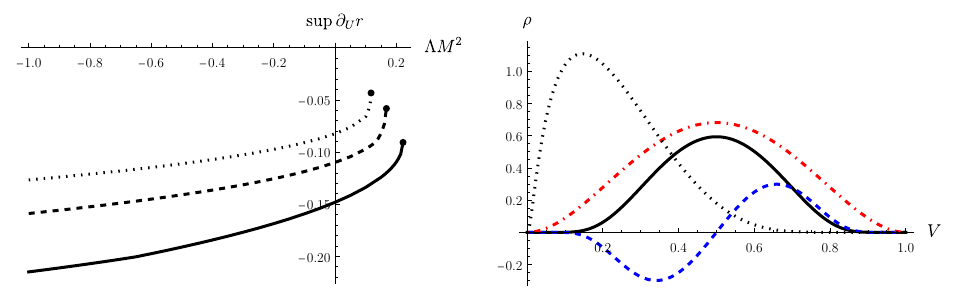}
    \caption{\textit{Left:} $\sup\partial_Ur$ (in units $M=1$) against $\Lambda M^2$ for the gluing solution achieved with the minimal value of $\e M$, with $\m=0$. The different curves represent different dimensionless charges: $\q=1$ (solid), $\q=0.9$ (dashed) and $\q=0.4$ (dotted). The dot denotes the maximum value of $\Lambda M^2$ (which depends on $\q$). In all cases we have $\sup \partial_U r<0$ so the candidate solution is a valid solution. \textit{Right:} Profiles $\rho$ for $\alpha_1=\alpha_{\rm max}$, for the even (solid black), odd (dashed blue), polynomial (dash-dotted red) and modified even (with $\gamma=0.36$; dotted black) Ans\"atze.}
    \label{fig:RN_C0_SupDr_Cosmo}
\end{figure}

Given a candidate solution we need to determine whether or not it is a valid solution by calculating $\sup \partial_U r$. In principle, the sign of this quantity depends on all of our dimensionless parameters $\q$, $\Lambda M^2$, $\e M$ and $\m/\e$. However, in practice we find that only the last of these affects this sign of $\sup \partial_U r$. With $\m/\e=0$ and fixing $\e M$ to the minimum value consistent with \eqref{eMbound}, we obtain the results shown in the first panel of Figure \ref{fig:RN_C0_SupDr_Cosmo} for different values of $\q$ and $\Lambda M^2$. In all cases the candidate solution satisfies the limiting condition $\sup \partial_U r<0$ and is therefore a valid solution. We will discuss what happens for $\m/\e \ne 0$ in Section \ref{sec:C0_massive} below.

Consider decreasing $\e M$ so that \eqref{eMbound} is saturated. In this limit we have $\alpha \rightarrow \alpha_{\rm max}$. The form of the scalar field profile for $\alpha=\alpha_{\rm max}$ is plotted in the second panel of Figure \ref{fig:RN_C0_SupDr_Cosmo} (the ``modified even'' Ansatz will be described in Section \ref{sec:C0_modified} below). For each Ansatz, this profile is the one that gives a candidate solution with prescribed values of $\q$ and $\Lambda M^2$ for the smallest possible value of $\e M$. Since equality is not permitted in \eqref{eMbound} this is a plot of the limiting form of the profile. In this limit, $\inf r \rightarrow 0$ on $\mathcal{C}$ and so candidate solutions close to achieving the limit are solutions for which the early time Minkowski (or (A)dS) region is very small.


In the limit $q \rightarrow 0$ at fixed $\e M$, \eqref{Ieq} gives $\alpha_1 \rightarrow 0$. This corresponds to gluing to a Schwarzschild-((A)dS) black hole. In this limit $r \rightarrow r_+$ throughout the gluing region so the scalar field is confined to a region in which $r$ does not vary significantly and the derivative of the scalar field w.r.t. $r$ becomes very large. This is a ``thin shell'' limit. One can instead glue to Schwarzschild with a smooth scalar field profile using an uncharged scalar field, i.e.~by taking $\e M \rightarrow 0$ as $\q \rightarrow 0$. 

\subsubsection{Massive scalar field}\label{sec:C0_massive}

As mentioned above, $\m$ does not enter the equations for a candidate solution so the only possible effect of non-zero $\m$ is to render invalid a candidate solution that is valid for $\m=0$. If $\m/\e$ is small enough then we find that this does not happen and so the results are identical to those for $\m=0$. 

\begin{figure}
    \centering
    \includegraphics[width=\linewidth]{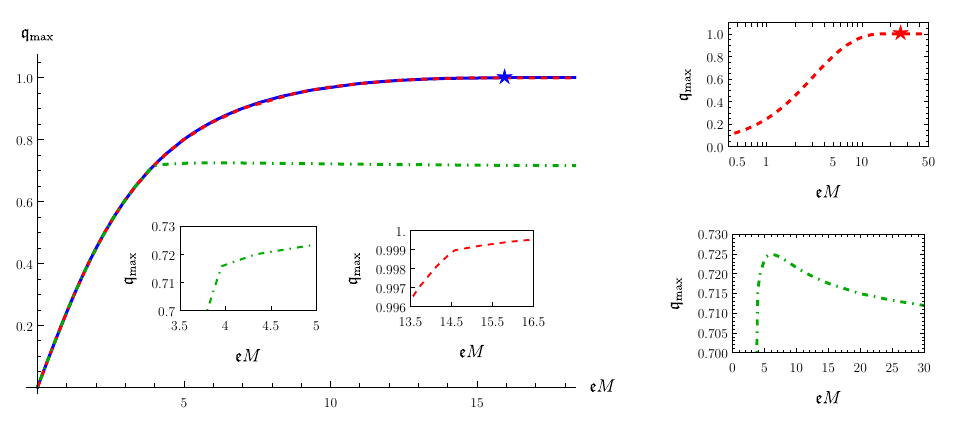}
    \caption{\textit{Left:} Maximum charge $\q_{\rm max}$ as a function of $\e M$ for $\m/\e=0$ (solid blue), $0.2$ (dashed red) and $0.5$ (dash-dotted green), with the even Ansatz. The blue curve attains extremality at $\e M\approx15.96$, denoted by a star. The latter two curves exhibit a kink when the limiting condition $\sup \partial_U r<0$ becomes important. The kink is more prominent for larger $\m/\e$; the insets show that of the red and green curves. After the kink, $\q_{\rm max}$ rises very slowly with $\e M$. The green curve reaches a maximum at $\q<1$, and then decreases slowly. \textit{Right, top:} $\q_{\rm max}$ versus $\e M$ with $\m/\e=0.2$, showing the very slow increase towards extremality, which is eventually reached at $\e M\approx25.97$, at the red star. Note the logarithmic scale on the horizontal axis. \textit{Right, bottom:} The slow rise and decrease in $\q_{\rm max}$ following the kink in the green curve ($\m/\e=0.5$). Note that all of this plot has much larger $\e M$ than the kink mentioned above.}
    \label{fig:RN_C0_mass_even}
\end{figure}

Once $\m/\e$ exceeds a critical value the results differ from $\m=0$. With zero cosmological constant, this critical value is $0.1876$, $0.1093$ and $0.2353$ for the even, odd and polynomial Ans\"atze, respectively. For $\m/\e$ above the critical value we obtain the results shown in Figure \ref{fig:RN_C0_mass_even} for the even Ansatz (results for the other Ans\"atze are qualitatively the same). For small $\e M$ the results are identical to the $\m=0$ results, with $\q_{\rm max}$ determined by equation \eqref{Ieq} with $\alpha_1=\alpha_{\rm max}$. However, once $\e M$ exceeds a certain value, candidate solutions with large $\alpha_1$ violate the limiting condition $\partial_U r<0$ and so are no longer valid solutions. This leads to a kink in the plot of $\q_{\rm max}$ against $\e M$. This shown in more detail in the inset of Figure \ref{fig:RN_C0_mass_even} (left). The kink occurs at lower $\e M$ for larger $\m/\e$. Beyond the kink, $\q_{\rm max}$ continues to rise, but much more slowly. For example, with the even Ansatz, when $\m/\e=0.2$, $\q_{\rm max}$ is identical to the $\m=0$ case up to the kink at $\e M\approx14.58$, but extremality is only reached much later, at $\e M\approx25.97$ (see top right panel of Figure \ref{fig:RN_C0_mass_even}), whereas the $\m=0$ case reaches extremality at $\e M\approx15.96$. If $\m/\e$ is too large, then extremality is never attained: $\q_{\rm max}$ eventually begins to decrease (extremely slowly) with larger $\e M$. This is shown in the bottom right panel of Figure \ref{fig:RN_C0_mass_even}.

This failure to attain extremality for large enough $\m/\e$ can be understood as follows. Notice that in \eqref{eq:EMCSF_waveR}, the squared mass is accompanied by $|\Phi|^2$. In order to keep $\partial_Ur$ negative everywhere, this term must be kept sufficiently small. The only way to keep this term small is to decrease the amplitude ($\propto\beta\equiv|\alpha_1|$) of the scalar field. Decreasing the amplitude requires increasing $\e M$ in order to achieve, say, $\q=1$. To determine the smallest value of $\e M$ for which extremality is possible, we must find the largest $\beta$ such that $\sup\partial_Ur<0$. As $\m/\e$ is increased, $\beta$ must decrease and $\e M$ increase. Eventually, $\beta$ approaches zero, and $\e M$ diverges to infinity. This implies that there is an upper bound on the value of $\m/\e$ for which one can reach extremality. We find that this bound is $0.2059$, $0.1114$ and $0.2773$ for the even, odd and polynomial Ans\"atze, respectively, with vanishing cosmological constant. These results are in agreement with \cite{Reall:2024njy}, which showed that an extremal Reissner-Nordstr\"om black hole cannot be formed in finite time in gravitational collapse if $\m/\e\geq1$.\footnote{This is perhaps surprising given that the proof of \cite{Reall:2024njy} doesn't hold for the low spacetime differentiability we are considering here. This is an example of how $C^0$ gluing already captures the most important features of $C^k$ gluing.}

When the cosmological constant is positive, the upper bound on $\m/\e$ increases slightly. For example, setting the cosmological constant to the maximum value of $\Lambda M^2=2/9$, with the polynomial Ansatz extremality is possible up to $\m/\e\approx0.3797$. This is an approximately $37\%$ rise in the upper bound compared with the asymptotically-flat case, but nevertheless remains very far from $\m/\e=1$. It is worth noting that reaching extremality with $\m>\e$ would not violate the results of \cite{Reall:2024njy}, as these apply only to asymptotically-flat spacetimes.\footnote{Note also the extension to this result for supersymmetric asymptotically-AdS black holes \cite{McSharry:2025iuz}, although this doesn't apply to the Reissner-Nordstr\"om-AdS black hole as it is not supersymmetric.}

\subsubsection{Modifying the Ansatz}\label{sec:C0_modified}

An interesting question is whether the minimum value of $\e M$ required to attain $\q=1$ can be made arbitrarily small by a suitable choice of Ansatz. A detailed analysis of this problem would require much more fine-tuning of the Ansatz and a complete investigation is beyond the scope of this paper. However, we can gain some understanding by introducing an additional parameter in our Ansatz. We will do so by defining a ``modified even Ansatz'', $\chi^m(V;\gamma)=\chi^e(V^\gamma)$, where $\chi^e(V)$ is the usual even Ansatz with a single bump ($K=1$ in \eqref{eq:even_Ansatz}).\footnote{We thank Christoph Kehle for suggesting this idea. Note that we are transforming the even Ansatz instead of the polynomial one (which requires lower $\e M$ to attain $\q=1$) because for $\gamma\leq 1/4$ the modified polynomial Ansatz would not give square integrable $\partial_V \Phi$, which, using \eqref{eq:EMCSF_RayV}, would be incompatible with existence of $\partial_V r$.} With multiple bumps---which will be required for more regular solutions later---the modified even Ansatz reads, in the notation of \eqref{eq:even_Ansatz},
\begin{equation}\label{eq:modified}
    \chi_j^m(V;\gamma)=N_m(K,\gamma)\, \exp\left[{\frac{1}{(V-(j-1)/K)^\gamma((V-(j-1)/K)^\gamma-1/K^\gamma)}}\right],\qquad \frac{j-1}{K}<V<\frac{j}{K}.
\end{equation}
The idea behind this Ansatz is that a factor of $\gamma^2$ appears on the right-hand side of the Raychaudhuri equation \eqref{eq:EMCSF_RayV}, while only a factor of $\gamma$ appears in the Maxwell equation \eqref{eq:EMCSF_MaxwellV}. Thus, choosing a small $\gamma$ might be expected to make the decrease in $r$ (arising when \eqref{eq:EMCSF_RayV} is solved backwards in $V$) parametrically smaller than the increase in $Q$ when \eqref{eq:EMCSF_MaxwellV} is solved forwards in $V$. Hence it should be easier to reach $\q=1$ without violating $r>0$. 

\begin{figure}
    \centering
    \includegraphics[width=0.5\linewidth]{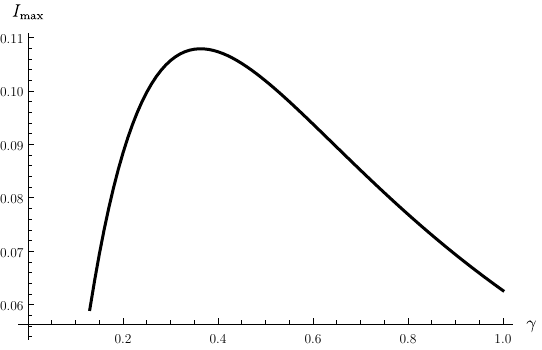}
    \caption{$I_{\rm max}$ versus $\gamma$ for the modified even Ansatz with parameter $\gamma$. The maximum is reached at $\gamma\approx0.36$.}
    \label{fig:RN_C0gamma}
\end{figure}

 We have investigated this Ansatz for $\m=0$. We find that the introduction of $\gamma$ indeed helps to reduce the minimum $\e M$ required to reach $\q=1$, but does not make it arbitrarily small. The smaller $\gamma$ is, the more the profile $\rho$ is peaked near $V=0$, and the sharper this peak is (see right plot of Figure \ref{fig:RN_C0_SupDr_Cosmo}). If we try to make $\gamma$ very small, the peak becomes too sharp, the right-hand side of the Raychaudhuri equation grows rapidly, and the amplitude of the scalar field must be decreased to ensure that $\inf r>0$. This, in turn, reduces $I_{\rm max}$. Our results, plotted in Figure \ref{fig:RN_C0gamma}, show that $I_{\rm max}$, as a function of $\gamma$, is a concave function with a maximum of about $0.1079$ at $\gamma=0.36$. The modified Ansatz favours a profile that is peaked closer to $V=0$---as long as it is not too sharp. 

The modified Ansatz with $\gamma=0.36$ peaks at $V\approx0.15$. If we reflect the Ansatz about $V=1/2$, the minimum value of $\e M$ required for extremality jumps by over a factor of $7$ to approximately $66.24$. This asymmetry can be explained as follows. Consider two profiles each involving a single compactly-supported bump. For one profile the bump has support at early time (near $V=0$) and for the other profile the support is at late time (near $V=1$). For either profile, equation \eqref{eq:EMCSF_RayV} implies that $\partial_V r$ is constant outside the support of the bump. When this equation is integrated backwards in time starting at $V=1$, $\partial_V r$ remains zero until it reaches the bump. After the bump it takes some constant positive value. The region in which $\partial_V r$ is non-zero is larger for the late-time bump. Hence, switching to a forward in time viewpoint, the increase in $r$ before the bump is larger for the late time bump. Hence the initial value of $r$ must be smaller for the late time bump. If we try to increase the injected charge by scaling up the amplitude of both bumps then the condition $\inf r>0$ will be violated for the late time bump before it is violated for the early time bump. For the example just discussed, the maximum amplitude of the optimally-parameterised modified Ansatz is nearly $4$ times larger than its reflected counterpart.

\subsubsection{Superextremal horizons}

The usual definitions of extremality, sub-extremality and super-extremality only make sense for stationary spacetimes (so only for $V\geq1$ in our situation). However, in spherical symmetry, we can formulate definitions applicable to a dynamical black hole using quasilocal mass and charge. For $\Lambda=0$ the \emph{renormalised Hawking mass} $\varpi$ is defined for a symmetry sphere of area-radius $r$ by
\begin{equation}
    1 - \frac{2\varpi}{r} + \frac{Q^2}{r^2} = g^{ab} (dr)_a (dr)_b
\end{equation}
In the Reissner-Nordstr\"om spacetime this gives $\varpi = M$ everywhere. This motivates the following definition in a dynamical spacetime, where $\varpi$ and $Q$ vary with $(U,V)$. We say that a symmetry sphere of given $(U,V)$ is subextremal if it has $\varpi>|Q|$, extremal if $\varpi  = |Q|$, and superextremal if $\varpi<|Q|$. In a dynamical black hole spacetime, we can apply this definition to a symmetry sphere on the event horizon to define what we mean by the horizon being extremal, subextremal or superextremal at a given value of $V$.

\begin{figure}
    \centering
    \includegraphics[width=\linewidth]{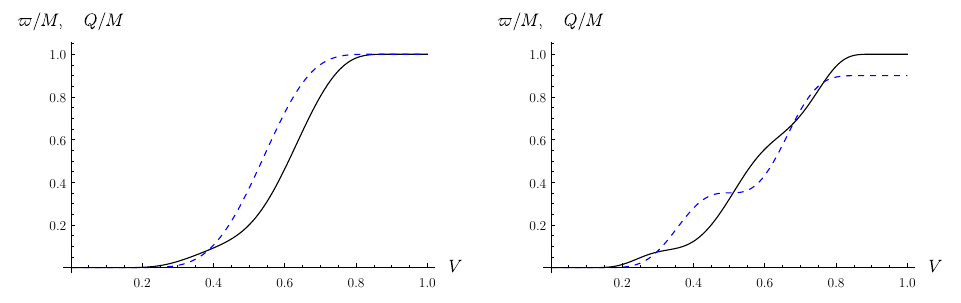}
    \caption{Renormalised Hawking mass $\varpi/M$ (black) and charge $Q/M$ (dashed blue) over time. \textit{Left:} Plots for a $C^0$ solution with the even Ansatz leading to the formation of an extremal black hole. \textit{Right:} Plots for a $C^0$ solution with the odd Ansatz leading to the formation of a black hole with $\q=0.9$. In both cases, the horizon is temporarily superextremal.}
    \label{fig:superextremal}
\end{figure}

For our solutions, the horizon must be sub-extremal or extremal for $V \ge 1$. However, we find that the horizon may be {\it super}extremal\footnote{We thank Christoph Kehle and Ryan Unger suggesting we investigate this.} for some interval (or union of intervals) of $V\in(0,1)$. It is only when the final Reissner-Nordstr\"om black hole is very far from extremality that the horizon remains subextremal for all times. For a solution that forms an extremal black hole, or a subextremal black hole of sufficiently high charge, we find that $Q>\varpi$ over a significant proportion of the full interval $V\in[0,1]$ (see Figure \ref{fig:superextremal}). This observation is independent of the Ansatz.\footnote{Similar results hold in the higher regularity problems.} At early time the horizon is always subextremal. However, often the charge increases faster, and thus eventually overtakes $\varpi$, bringing the horizon to a superextremal regime. If the Ansatz vanishes in the interior of the interval, then $\partial_VQ=0$ at those points, resulting in a series of ``steps'' in the curve of $Q$. This can lead to the horizon oscillating between subextremal and superextremal phases. An example is shown in the right panel of Figure \ref{fig:superextremal}. At late times, the renormalised Hawking mass increases faster than the charge, and eventually becomes equal to it (if $\q=1$) or larger than it (if $\q<1$). In the former case, the horizon is superextremal immediately before it becomes extremal.

\subsection{$C^1$ Gluing}\label{sec:C1}

$C^1$ gluing is performed in a similar manner to $C^0$ gluing. The differences are that our Ans\"atze each contain $3$ real parameters $\alpha_i$ and, in addition to the steps required for $C^0$ gluing, we now also need to glue the complex quantity $\partial_U \Phi$. To do this we solve \eqref{eq:EMCSF_wavePhi} on $\mathcal{C}$ with initial condition $\partial_U \Phi(0)=0$ and try to choose the parameters so that $\partial_U \Phi(1)=0$ along with \eqref{Ieq}. If we can do this, we have a candidate solution, and for this to be a valid solution the limiting condition $\partial_U r<0$ must be satisfied, as for the $C^0$ case.

\subsubsection{Restricted parameter space and $\mathfrak{Q}^2$}

\begin{figure}
    \centering
    \includegraphics[width=\linewidth]{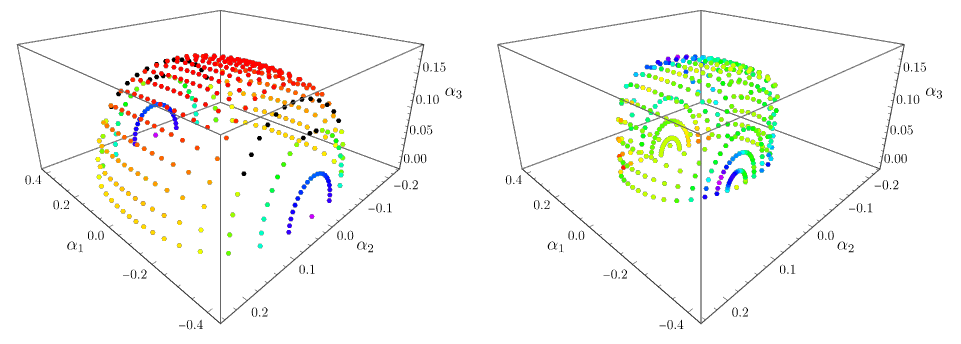}
 \caption{\textit{Left:} Boundary of the restricted parameter space for the even Ansatz. Only one half is shown; the other can be obtained by a reflection through the origin. Colours indicate the value of $I$. Large values are in bluer tones, and tend to accumulate in the region where $\alpha\approx(\pm\beta,0,0)$, corresponding to a scalar field profile for which the first compactly supported pulse has much larger amplitude than the two subsequent pulses. The minimum value of $I$ on the boundary is $0.0009$ and the maximum is $0.0041$; the latter is also the global maximum. Black points denote boundary points where $\partial_\beta I<0$, corresponding to directions $\hat{\alpha}$ for which the maximum of $I$ along a ray with direction $\hat{\alpha}$ is not attained at the boundary. \textit{Right:} One half of the set $\mathfrak{Q}^2$ for the even Ansatz with $I\approx0.0009$, the largest value such that $\mathfrak{Q}^2$ is topologically spherical, touching the boundary of the restricted parameter space at the North pole. The colours indicate the magnitude of $\partial_U\Phi(1)$ for $\q=1$ and $\Lambda=\m=0$, with reddish tones (located on the left side) corresponding to $\alpha$ near a candidate solution.}
    \label{fig:RN_C1_restricted}
\end{figure}

We find that the restricted parameter space has the topology of a ball in $\mathbb{R}^3$. The left panel of Figure \ref{fig:RN_C1_restricted} shows the boundary of this ball for the even Ansatz, colour coded according to the value of $I$ on the boundary. For most directions $\hat{\alpha}$, the maximum of $I$ occurs at the boundary of the restricted parameter space. However, there are some directions (corresponding to the black points) for which the maximum occurs at an interior point. Nevertheless, the global maximum of $I$ occurs on the boundary. Recall that larger $I$ corresponds to smaller $\e M$ (for given $\q$ and $\Lambda M^2$) for candidate solutions of the $C^0$ gluing problem. 

Next we consider the surface $\mathfrak{Q}^2$ corresponding to parameter values that achieve the desired value of $\q$ for given $\Lambda M^2$ and $\e M$, i.e.~the surface of constant $I$ defined by $\alpha$ satisfying equation \eqref{Ieq}. For small $Q/(\e r_+^2)$, $\mathfrak{Q}^2$ is a topological sphere centered at the origin \cite{Kehle:2022uvc}. As $Q/(\e r_+^2)$ increases, we find that the size of the sphere increases until it touches the boundary of the restricted parameter space at the North pole (i.e.~the point $\alpha=(0,0,\beta)$). This occurs for $Q/(\e r_+^2)\approx0.0009$, corresponding to the surface shown on the right plot of Figure \ref{fig:RN_C1_restricted}. At still larger $I$ a hole appears in $\mathfrak{Q}^2$, so its topology is no longer spherical. If $Q/(\e r_+^2)$ is increased even further then eventually all of $\mathfrak{Q}^2$ will reach the boundary and solutions of \eqref{Ieq} will no longer exist. The qualitative behaviour of the surface $\mathfrak{Q}^2$ is the same for all of our Ans\"atze. 

Note that our Ans\"atze for $C^0$ gluing are not special cases of our Ans\"atze for $C^1$ gluing so we cannot compare directly our results for $C^0$ gluing to our results for $C^1$ gluing. However, we can instead use a $C^1$ Ansatz to perform $C^0$ gluing. Each point of $\mathfrak{Q}^2$ then corresponds to a candidate solution for $C^0$ gluing and the maximum value of $I$ then dictates the maximum value of $Q/(\e r_+^2)$ for which $C^0$ gluing is possible, equivalently the minimum value of $\e M$ for which $C^0$ gluing is possible for given $\q$ and $\Lambda M^2$. This maximum value gives $1/I_{\rm max} \approx 244$, $765$ and $7.0$ for the even, odd and polynomial Ans\"atze, respectively. For $\Lambda=0$ and $\q=1$ this is the minimum value of $\e M$ for which a candidate solution for $C^0$ gluing exists with this Ansatz. The limiting condition $\sup\partial_Ur$ behaves in a qualitatively identical way to that in $C^0$ gluing, as in the first panel of Figure \ref{fig:RN_C0_SupDr_Cosmo}. In particular, whenever $\m=0$, we have $\sup\partial_Ur<0$ for all points in the restricted parameter space. Thus all points of $\mathfrak{Q}^2$ correspond to valid solutions of $C^0$ gluing.

\subsubsection{Massless scalar field}

The colour on the right plot of Figure \ref{fig:RN_C1_restricted} shows how close each point is to satisfying the gluing condition for $\partial_U \Phi$, with $\q=1$ and $\Lambda=\m=0$. In this example the gluing condition is satisfied by a single point (and its image under $\alpha \rightarrow - \alpha$), lying in the red region. Hence there exists a single candidate solution in this case (modulo the reflection symmetry). This is typical: we have not found any examples of $C^1$ gluing for which there exist multiple solutions (with a massless scalar field). In our example, the red region does not lie near the black dots at the boundary of the restricted parameter space. This implies that, as we decrease $\e M$, a candidate solution continues to exist until the red region reaches the boundary of the restricted parameter space. For smaller $\e M$ no candidate solution exists. As mentioned above, with a massless scalar field we find that $\sup\partial_Ur<0$ everywhere in the restricted parameter space, so all candidate solutions are also valid solutions of $C^1$ gluing. 


All of our Ans\"atze exhibit qualitatively the same behaviour. For $\q=1$ and $\Lambda=\m=0$ we find that the minimum value of $\e M$ for which a valid solution of $C^1$ gluing exists is $(\e M)_{\rm min}=720$, $2909$, $78.5$ and $239$ for the even, odd, polynomial and modified Ans\"atze respectively. The optimal parameter for the modified Ansatz is now $\gamma\approx0.48$. As would be expected, these are somewhat larger than the minimum values required when a $C^1$ Ansatz is used for $C^0$ gluing, as determined above. Notice that the polynomial Ansatz has a less constraining bound on $\e M$ than the modified Ansatz, unlike in the $C^0$ case. This is because the latter Ansatz now is formed of three bumps, which increases the size of the derivative term $|\partial_V\Phi|^2$ in the Raychaudhuri equation \eqref{eq:EMCSF_RayV} but adds little charge to the system.

\begin{figure}
    \centering
    \includegraphics[width=\linewidth]{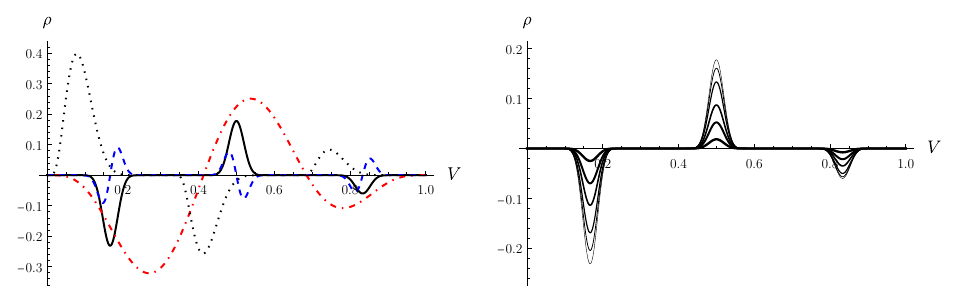}
    \caption{\textit{Left:} Profiles $\rho$ providing solutions to the $C^1$ gluing problem for the even (solid black), odd (dashed blue), polynomial (dash-dotted red) and modified ($\gamma=0.48$, dotted black) Ans\"atze for $\q=1$ and $\Lambda=\m=0$. The value of $\e M$ used was (approximately) the smallest that allowed $\q=1$ to be reached, namely $720$, $2909$, $78.5$ and $239$, respectively. \textit{Right:} Profiles for the even Ansatz with $\e M=720$, for $\q=0.1$, $0.6$, $0.9$, $0.99$, $0.999$ and $1$ (thickest to thinnest).}
    \label{fig:RN_phi_solsC1}
\end{figure}

In Figure \ref{fig:RN_phi_solsC1} we plot examples of scalar field profiles for the $C^1$ problem. The first panel shows profiles that glue to an extremal black hole with the smallest possible $\e M$, for each Ansatz. As in Section \ref{sec:C0}, we find that profiles that are larger in amplitude near the Minkowski region, where $r$ is smaller, and lower nearer $V=1$, are favoured. As $\q$ is varied for constant $\e M$, we find, as expected, that the amplitude of the profile must grow with $\q$. This is illustrated in the second panel of Figure \ref{fig:RN_phi_solsC1}. In the case of the even, odd and modified Ans\"atze, for example, each bump's amplitude increases, although not all by the same amount, i.e.~in the notation of Section \ref{sec:gen_approach}, $\hat{\alpha}$ also changes, but only slightly. The same is true for the polynomial Ansatz: the shape remains qualitatively similar over all $\q$, and it is mostly the amplitude that changes. Similarly, $\hat{\alpha}$ changes only very slightly when $\Lambda M^2$ is varied. To quantify what we mean by ``$\hat{\alpha}$ changes slightly'', it is useful to parametrise $\hat{\alpha}$ with angles $(\theta,\phi)$ on $S^2$. Fixing $\e M$, $\Lambda M^2$ and the Ansatz, we find that as $\q$ is varied from (nearly) zero to $\q_{\rm max}$, the angles of the candidate solution change by an amount on the order of $0.1$, or smaller. By comparison, note that the domain of $\theta$ is $[0,\pi]$ and that of $\phi$ is $[0,2\pi)$. Similarly, when the Ansatz is fixed and $\q$ maximised for given $\e M$, then as $\e M$ is varied each angle again changes by an amount on the order of $0.1$, or less. Finally, if we fix $\q$ and $\e M$ and the Ansatz, and vary $\Lambda M^2$, the change is very small: for example, between $\Lambda M^2=-1$ and $\Lambda M^2=2/9$ the angles differ by an amount on the order of $0.01$ or less. This is true for a wide range of $\q$ and $\e M$. The result that $\hat{\alpha}$ varies only slightly as the parameters vary implies that saturating the inequality \eqref{eMbound} can be used to give an approximate result for $\q_{\rm max}$ (the maximum value of $\q$ achievable for given $\e M$ and $\Lambda M^2$) if $I_{\rm max}$ is interpreted as the maximum value of $I$ in the direction $\hat{\alpha}$. The second panel of Figure \ref{fig:RN_C0_I} then provides a good approximation of how $\q_{\rm max}$ varies as a function of $\e M$ and $\Lambda M^2$.

It is interesting to compare the minimum value of $\e M$ for which $\mathfrak{Q}^2$ is smooth, as predicted by \cite{Kehle:2022uvc} for large $\e M$, with the (lower) minimum value of $\e M$ for which a valid solution exists. This comparison gives a rough measure of how much our numerical approach goes beyond the regime of validity of the proof of \cite{Kehle:2022uvc}.
We have investigated this for the extremal case $\q=1$ (again with $\Lambda=\m=0$). For the even Ansatz, the former bound is $\e M\gtrsim 1141$, but a valid solution exists for $\e M\gtrsim 720$: a decrease of nearly $37\%$. For the odd Ansatz, these bounds are $3531$ and $2909$, respectively, representing a more modest reduction of approximately $18\%$. For the polynomial Ansatz the bounds are $90.0$ and $78.5$ respectively, a decrease of $13\%$ (although the polynomial Ansatz always lies outside the regime of validity of \cite{Kehle:2022uvc} since the latter considers only disjoint compactly-supported functions $\chi_i$).

\subsubsection{Massive scalar field}\label{sec:mass}

As in the $C^0$ problem, the main effect of a non-zero scalar mass is to increase $\sup\partial_Ur$, thereby making it more likely to take the incorrect sign. However, now $\m$ also affects $\partial_U\Phi$ from \eqref{eq:EMCSF_wavePhi} and hence has an effect on candidate solutions. Thus, when $\m/\e$ is varied (with the other parameters fixed), $\alpha$ must be adjusted in order to obtain a candidate solution, which in turn affects $\inf r$, leading to the possibility of the parameters reaching the boundary of the restricted parameter space. We will see that this, in part, leads to different behaviours depending on the Ansatz. The results described below are all for $\Lambda=0$.

\begin{figure}
    \centering
    \includegraphics[width=\linewidth]{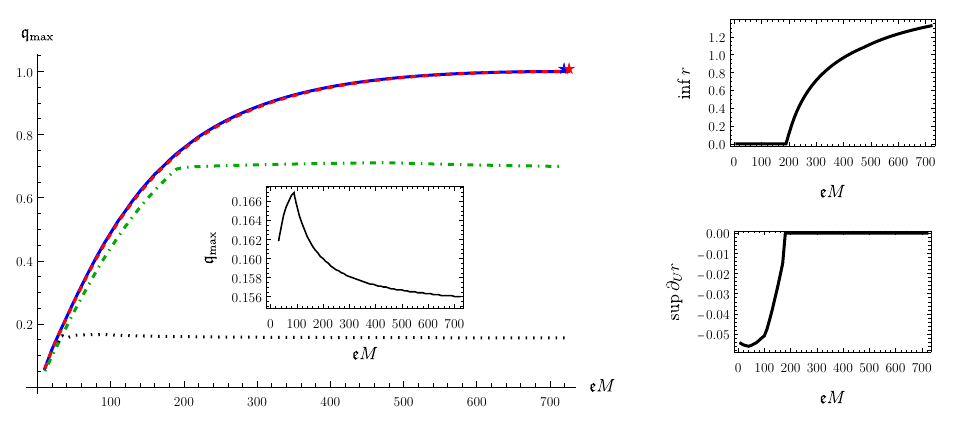}
    \caption{$C^1$ gluing with a massive scalar field.  \textit{Left:} Maximum charge $\q_{\rm max}$ as a function of $\e M$ for $\Lambda=0$ and $\m/\e=0$ (solid blue), $0.01$ (dashed red, lying nearly on top of the blue curve), $0.07$ (dash-dotted green) and $0.35$ (dotted black). The former two curves attain extremality at $\e M\approx720$ and $727$, respectively, denoted by stars. The latter two curves exhibit a kink when the limiting condition $\sup \partial_U r$ becomes important. After the kink, the maximum charge rises very slowly with $\e M$, reaches a maximum, and decreases slowly. The inset zooms in on this phenomenon for the case of $\m/\e=0.35$. The peak in the inset is not the kink but is instead part of the near-flat segment. \textit{Right:} $\inf r$ (top) and $\sup\partial_Ur$ (bottom) as a function of $\e M$ for the gluing solution with $\q=\q_{\rm max}(\e M)$, with $\m/\e=0.07$.}
    \label{fig:RN_mass_comp}
\end{figure}

For the even Ansatz, when $\m/\e$ is increased with $\q$ and $\e M$ fixed, $\inf r$ decreases so the solution moves closer to the boundary of the restricted parameter space. This implies that even for very small $\m/\e$, the lower bound on $\e M$ to attain some given $\q$ grows with $\m/\e$. For instance, the minimum value of $\e M$ required to attain extremality grows from about $720$ at $\m/\e=0$ to approximately $786$ at $\m/\e=0.0265$. In those cases it is still $\inf r$ that sets the bound on $\e M$. In the $C^0$ case the bounds would be identical, as $\inf r$ would not be affected. For higher $\m/\e$, the curve of $\q_{\rm max}$ versus $\e M$ experiences a kink, much like in the $C^0$ case, beyond which the limiting condition $\sup\partial_Ur<0$ becomes important. Sample curves are provided in Figure \ref{fig:RN_mass_comp}. The right panels show how $\inf r$ and $\sup \partial_U r$ vary for the solution with the largest value of $\q$ for which gluing can be achieved. This shows how $\q_{\rm max}$ is determined by $\inf r>0$ before the kink and $\sup \partial_U r<0$ after the kink. Beyond the kink, $\q_{\rm max}$ grows very slowly with $\e M$, and, if $\m/\e\gtrsim0.02720$, fails to reach extremality. Thus the behaviour of the solutions looks almost qualitatively identical to Figure \ref{fig:RN_C0_mass_even}, except that the curves for different $\m/\e$ deviate slightly even for small $\q_{\rm max}$.

\begin{figure}
    \centering
    \includegraphics[width=\linewidth]{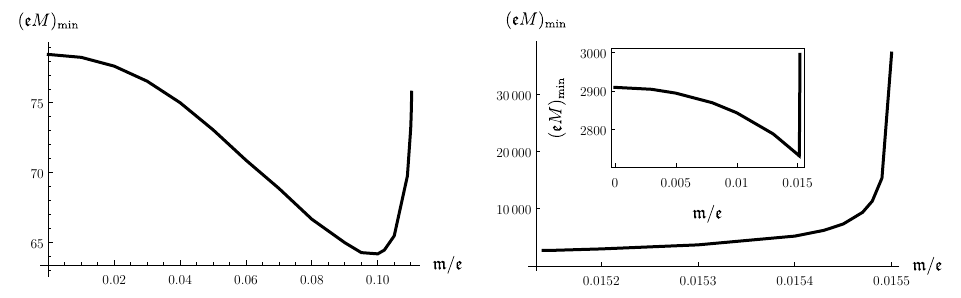}
    \caption{\textit{Left:} Minimum value of $\e M$ required to reach extremality with the polynomial Ansatz as a function of $\m/\e$. The curve diverges for sufficicntly large $\m/\e$. \textit{Right:} Same data but for the odd Ansatz. The main plot shows only the segment where $\sup\partial_Ur$ is the limiting condition. Here, $\e M$ appears to blow up as $\m/\e$ is increased. The inset shows the data for lower mass-to-charge ratios. When $\inf r$ is the limiting condition, the curve decreases monotonically, but then exhibits a sharp kink when $\sup\partial_Ur$ takes over.}
    \label{fig:RN_mass_poly}
\end{figure}

Surprisingly, the odd and polynomial Ans\"atze behave rather differently. When $\m/\e$ is increased, the minimal $\e M$ necessary to obtain extremality at first \emph{decreases}, as shown in Figure \ref{fig:RN_mass_poly}, in stark contrast to the case of the even Ansatz (for which the corresponding plot would just increase monotonically, and diverge at $\m/\e = 0.02720$). This is because, for $\q=1$ and fixed $\e M$, $\inf r$ increases with $\m/\e$, unlike for the even Ansatz. Hence one can slightly lower $\e M$ and still form an extremal black hole, i.e.~for small $\m/\e$, increasing $\m/\e$ makes it slightly easier to achieve extremality. This is intriguing, as we know from \cite{Reall:2024njy} that when $\m/\e\geq 1$, $\q=1$ is impossible. So, for large enough $\m/\e$ we would expect the minimum value of $\e M$ to blow up to infinity as $\q\to1$, or even earlier, as we see in the case of the even Ansatz. It turns out that the bound of \cite{Reall:2024njy} is still satisfied, as is shown in Figure \ref{fig:RN_mass_poly}.

Suppose we increase $\m/\e$, while always taking $\e M$ to be the lowest value that permits extremality, $\q=1$. During this process, the gluing solution sees $\sup\partial_Ur$ approach zero, although initially it is still $\inf r$ that determines $\q_{\rm max}$. Thus, at first $\e M$ decreases as $\m/\e$ is raised. However, when the limiting condition $\sup\partial_Ur<0$ becomes relevant, then as $\m/\e$ is pushed further up, $\e M$ must be increased to obtain extremality. For the odd Ansatz, this leads to the sharp kink visible in the right panel of Figure \ref{fig:RN_mass_poly}. In contrast to both the even and odd Ans\"atze, we have not identified any kinks in the curve of the polynomial Ansatz. In all cases, the mass-to-charge ratio cannot be increased indefinitely: beyond some critical value, valid solutions fail to exist, as $(\e M)_{\rm min}$ blows up to infinity as $\m/\e$ increases. For the polynomial Ansatz, this occurs at $\m/\e\approx0.1103$, whereas for the odd Ansatz it occurs at $\m/\e\approx0.01550$.

Given the range of behaviours observed with different Ans\"atze, and the fact that these results depend on the regularity class, it is difficult to draw general conclusions about characteristic gluing with a massive scalar field. For small $\m/\e$, the condition determining the minimum value of $\e M$ required to achieve a given $\q$ is $\inf r>0$, and this quantity can either decrease or increase as $\m/\e$ is varied (with $\e M$ and $\q$ fixed). In all cases, when the limiting condition $\sup\partial_Ur <0$ becomes important, a change in behaviour is induced, although the details of this change depends on the Ansatz. Nevertheless, there is always an upper bound on the value of $\m/\e$ for which extremality can be achieved, and this bound is in agreement with \cite{Reall:2024njy}.

\subsection{$C^2$ Gluing}\label{sec:C2}

We now study the $C^2$ gluing problem, in which we additionally constrain the second transverse derivative of the scalar field to vanish at $V=1$. Thus, the profile of the scalar field must depend on five parameters. Unfortunately, searching for solutions in a $5$-dimensional space is significantly more complicated than in a $3$-dimensional space, and our numerical methods have proved far less reliable than in the $C^1$ case. Moreover, we cannot easily visualise the search space, so identifying suitable initial guesses is difficult. As a result, below we only consider the even Ansatz, for which we have obtained sufficient data to draw some conclusions.

Many of the results of the $C^0$ and $C^1$ gluing problems follow through to the $C^2$ case. There is an important difference, however: for a given set of parameters $(\q,\e M,\m/\e,\Lambda M^2)$, there exist multiple solutions $\alpha$. When one of the parameters is slowly varied with the others kept fixed, each solution changes continuously, and we can trace the solutions as functions of that parameter. We will refer to these traces as \emph{branches}.

\begin{figure}
    \centering
    \includegraphics[width=\linewidth]{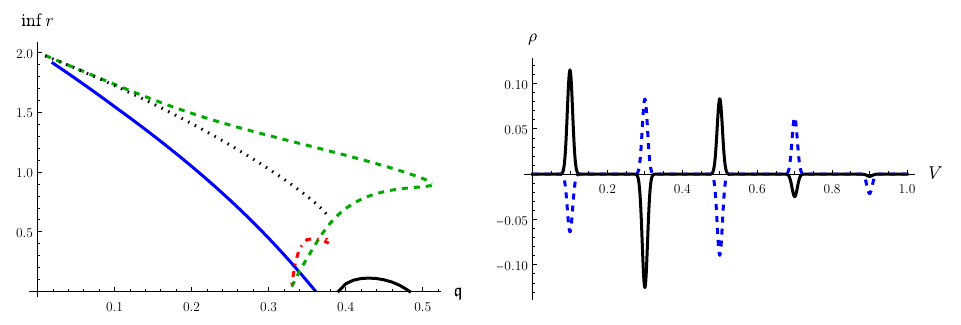}
    \caption{\textit{Left:} $\inf r$ versus $\q$ of various solution branches at $\e M=600$, $\m/\e=0$, $\Lambda M^2=0$. It is likely that there exist more solutions than presented here. \textit{Right:} Profiles for $C^2$ solutions which form an extremal black hole at $\e M=4417$ (black) and $\e M=6198$ (blue), with $\m=0=\Lambda$.}
    \label{fig:RN_C2_profiles}
\end{figure}

To study how the various branches behave, it is easiest to first try to find as many solutions as possible with fixed parameters. The left panel of Figure \ref{fig:RN_C2_profiles} shows a number of solutions obtained with $\e M=600$, $\m/\e=0$, $\Lambda M^2=0$, and also shows how $\inf r$ varies with $\q$ for each of them. The picture is rather complicated; branches can cease existing due to $\inf r$ becoming negative, or they can sometimes merge with each other, thereby annihilating themselves.\footnote{Annihilation via merging is a well-known phenomenon in bifurcation theory. For instance, consider the polynomial $f(x;s)=x^2+s$ depending on some parameter $s$. For $s<0$ there are two real roots. As $s\to0-$, the two roots converge, and become one multiple root. There are no real roots for $s>0$.} Due to the complicated structure, our numerical algorithm is sometimes unstable, and thus there may well exist solutions we have not found. For instance, it seems likely that the branches that disappear without crossing $\inf r=0$ merge with another solution branch which---apart from the dashed green branch in Figure \ref{fig:RN_C2_profiles}---our algorithm has not converged to.

Fixing $\m=0$, the branches that do not merge with another seem to behave in the same way as the unique branches of the $C^0$ and $C^1$ gluing problems. In particular, for each branch, $\sup\partial_Ur<0$, for any value of $\Lambda M^2$. Moreover, as in the $C^1$ case, along each branch $\hat{\alpha}$ does not vary significantly with $\q$ and $\Lambda M^2$, so the second panel of Figure \ref{fig:RN_C0_I} gives a good approximation of how the maximum attainable charge depends on $\e M$ and $\Lambda M^2$. (The variance in $\hat{\alpha}$ is roughly similar to that of the $C^1$ case.) The minimum value of $\e M$ needed to attain extremality is different for each branch: approximately $4417$ and $6198$ for the black and blue branches (see Figure \ref{fig:RN_C2_profiles}) respectively, with $\Lambda=0$.

Representative profiles of the two branches we have found that attain extremality are shown in the right panel of Figure \ref{fig:RN_C2_profiles}. The profiles chosen are those which achieve extremality with the minimum value of $\e M$. As in $C^0$ and $C^1$ gluing, the profiles reach their maximum amplitude in the first half of $\mathcal{C}$, i.e.~for $V<1/2$. Note however that the amplitudes of each peak do not decrease monotonically.

\begin{figure}[t]
    \centering
    \includegraphics[width=\linewidth]{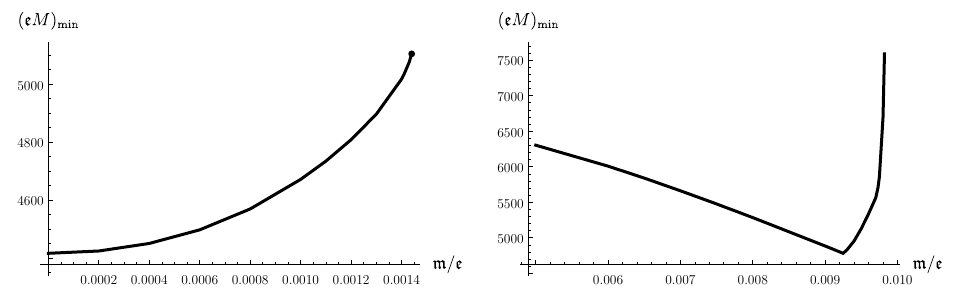}
    \caption{Minimum value $(\e M)_{\rm min}$ necessary to reach extremality as a function of $\m/\e$, for two branches of the even Ansatz in the $C^2$ problem. Here $\Lambda=0$. \textit{Left:} The branch ceases to exist at the point. Beyond, we find $\inf r<0$. \textit{Right:} Here the branch initially has decreasing $(\e M)_{\rm min}$ but eventually diverges to infinity. The sudden increase arises due to $\sup\partial_Ur$ approaching zero.}
    \label{fig:RN_C2_mass_cosmo}
\end{figure}

Let us now consider non-zero $\m$. As $\m/\e$ is varied, different branches change in different ways---even though these branches are all constructed from the same basic profile. Two examples are shown in Figure \ref{fig:RN_C2_mass_cosmo}. In the left panel, the minimum value of $\e M$ needed to form an extremal black hole increases with $\m/\e$. However, this process stops early, as for $\m/\e\gtrsim0.00144$, we cannot find any candidate solutions (in particular, $\inf r>0$ fails). The right panel shows another example, where the behaviour is more akin to that we observe in the $C^1$ problem. For low mass-to-charge ratios of the scalar field, $(\e M)_{\rm min}$ decreases with $\m/\e$, but this relationship turns around sharply when $\sup\partial_Ur\to0-$, and $(\e M)_{\rm min}$ then diverges to infinity. The maximum $\m/\e$ of the scalar field with which one can achieve $\q=1$ is approximately $0.00992$. Much like in the lower regularity problems, it is difficult to draw sweeping conclusions about gluing with a massive scalar field, as different branches can behave in very different ways.

\section{Discussion}\label{sec:discussion}

The gluing solutions are strongly dependent on the Ansatz used for the scalar field profile along the null cone. For a given Ansatz, extremality can be reached only if $\e M$ is sufficiently large, and $\m/\e$ sufficiently small. We have obtained lower and upper bounds, respectively, for these parameters, for several Ans\"atze, up to the $C^2$ gluing problem. Our results are summarised in Table \ref{tab:results}. When the scalar field is massless, the lower bound on $\e M$ is determined by the requirement that $\inf r>0$. This is true for all Ans\"atze we have studied, for any value of the cosmological constant. As shown in Table \ref{tab:results}, a positive cosmological constant reduces the lower bound on $\e M$. A negative cosmological constant instead raises this bound. The bound is well approximated by $\e M  r_+^2/Q M={\rm constant}$ (this expression is exact for $C^0$ gluing). The sign and size of $\Lambda M^2$ does not qualitatively change the remaining properties of the gluing solutions compared with the $\Lambda=0$ case. Gluing solutions of different regularity and derived from different Ans\"atze generally behave in the same (qualitative) way. (In the $C^2$ problem, some solution branches behave differently, but these typically cannot form an extremal black hole.) The exception is when the scalar field is massive. In this case, for $C^1$ and higher gluing, different Ans\"atze and different branches of the same Ansatz may behave in different ways. The main difference between $C^1$ and $C^2$ gluing is the presence of additional branches in the latter case.

The results of Table \ref{tab:results} suggest that, for $C^k$ gluing, the lower bound on $\e M$ required to reach extremality increases very rapidly with $k$. However, as we discussed in Section \ref{sec:C0}, this is mostly a consequence of the fact that our Ansatz for $C^0$ gluing (say) is not a special case of our Ansatz for $C^1$ gluing. If we instead use the $C^1$ Ansatz to perform $C^0$ gluing then the increase from the minimum value of $\e M$ required for $C^0$ gluing to the minimum value required for $C^1$ gluing is much smaller than in Table \ref{tab:results}, as discussed in Section \ref{sec:C0}. 

We expect that the results for $C^0$, $C^1$ and $C^2$ gluing broadly follow through to higher regularity classes too. It is likely that these will involve multiple branches. We expect that these will behave similarly to branches in the lower regularity problems. This motivates the following question: is it only in the Einstein-Maxwell-charged scalar field model that low-regularity gluing qualitatively captures the features of gluing with higher regularity, or is this a generic feature of characteristic gluing? An answer in the affirmative would be very interesting, as it is significantly easier to study the $C^0$ and $C^1$ cases, even compared to only $C^2$. Note that a classical ($C^2$) solution to the Einstein equations in the \emph{entire} spacetime requires $C^3$ characteristic data, as the wave equation loses one derivative at the centre $r=0$ \cite{Kehle:2022uvc}. Proving that an analysis of $C^0$ gluing, or perhaps more likely $C^1$ gluing, is sufficient to obtain a qualitative understanding of the $C^3$ problem would therefore be very beneficial.

One may wonder why the $C^2$ problem can have multiple solutions for a given $(\q,\e M,\m/\e,\Lambda M^2)$, whereas the $C^1$ case only has one. Since the latter has fewer conditions to satisfy, these observations appear counterintuitive. However, recall that we are choosing our Ansatz to have as many free parameters as there are equations to be solved. While $C^2$ gluing requires two more constraints, we are searching for solutions in a $5$-dimensional space instead of a $3$-dimensional space, so perhaps it is not unreasonable to find more solutions. Of course, there is no physical reason to restrict the number of free parameters in the Ansatz to a number depending on the regularity we desire. For example, as discussed above, we can use the $C^1$ Ansatz to obtain infinitely many solutions of $C^0$ gluing.

One goal of this paper was to determine the minimum value of $\e M$ required to achieve extremality for a given Ansatz (for, say, $\Lambda=\m=0$). Our numerical methods have shown that, for any given Ansatz, this minimum value is significantly lower than the (large) value required in the proof of \cite{Kehle:2022uvc}. Of course, the most interesting question is what is the minimum value of $\e M$ when we extremize over all possible Ans\"atze. In particular, is this minimum value zero, in other words, for any fixed $\e$ is it possible to form an arbitrarily small extremal black hole in finite time?\footnote{We thank Christoph Kehle for raising this question.} Or if the minimum value is non-zero then does the answer depend on $k$? This is clearly a more difficult problem than the one we have tackled. But maybe it can be investigated numerically by devising a method to minimize $\e M$ for an Ansatz depending on a large number of parameters. 

A similar question concerns the massive scalar field: if we extremize over all possible Ans\"atze then what is the minimum value of $\m/\e$ required to achieve extremality? The largest values of $\m/\e$ for which we have managed to glue to extremal Reissner-Nordstr\"om are all well below $1$ so our results are consistent with the proof of \cite{Reall:2024njy} that gravitational collapse to form extremal Reissner-Nordstr\"om in finite time is not possible if $\m/\e \ge 1$. Perhaps by investigating Ans\"atze with a large number of parameters it will be possible to explore whether or not this bound is sharp.

All of our Ans\"atze have the same phase $e^{-iV}$, which was motivated by the proof of \cite{Kehle:2022uvc}. It would be interesting to explore the effects of making different choices for the phase function. The extra freedom in the phase may also help to minimise the value of $\e M$ and maximise the value of $\m/\e$ at which gluing to an extremal black hole is possible.

\begin{table}[t]
    \centering
    \begin{tabular}{ccc|cccc}
         & & $\Lambda M^2$ & Even & Odd & Polynomial & Modified \\
         \hline\hline
        $C^0$ & $\min \e M$ & 0 & 15.96 & 53.86 & 10.71 & 9.27\\ 
        & $\min \e M$ & 2/9 & 7.53 & 25.40 & 5.05 & 4.37 \\ 
        & $\max \m/\e$ & 0 & 0.2059 & 0.1114 & 0.2773 & \\ \hline
        $C^1$ & $\min \e M$ & 0 & 720 & 2909 & 78.5 & 239\\
        & $\min \e M$ & $2/9$ & 339 & 1372 & 37.2 & 112\\
         & $\max \m/\e$ & 0  & 0.02720 & 0.01550 & 0.1103 & \\ \hline
        $C^2$ & $\min \e M$ & 0  & 4417 & & &  \\
        & $\min \e M$ & $2/9$  & 2076 & & &  \\
        & $\max \m/\e$ & 0  & 0.00992 & & &
    \end{tabular}
    \caption{Bounds on $\e M$ and $\m/\e$ to reach extremality for different Ans\"atze.}
    \label{tab:results}
\end{table}

Another objective of this paper was to give concrete examples of gluing solutions produced by the construction of \cite{Kehle:2022uvc}. Having achieved this, the next step would be to give concrete examples of the resulting spacetime solutions. This could be done by using numerical methods to extend our gluing solutions away from the horizon. This would provide insights into the properties of Cauchy data sets whose maximal developments lead to violations of the third law. This might shed light on a question discussed in \cite{Dafermos:2025int}, namely whether initial data leading to the formation, in finite time, of an extremal Reissner-Nordstr\"om black hole is in any sense ``more fined tuned'' than initial data leading to the formation, in finite time, of a Reissner-Nordstr\"om black hole with some prescribed subextremal charge to mass ratio.

Finally, it is interesting to ask whether the results presented in this paper and in \cite{Kehle:2022uvc} have analogues for black holes in vacuum. Kehle and Unger conjecture that it is indeed possible to form an extremal Kerr black hole in vacuum, in finite time. If true, such a result would give a disproof of the third law that is independent of any choice of matter model. However, there are (at least) two reasons why investigating this using gluing is a much harder problem than the problem of gluing to extremal Reissner-Nordstr\"om. First, we no longer have spherical symmetry: in vacuum the free data for the gluing problem is the shear of the null generators of $\mathcal{C}$, and this is now a function of $3$ coordinates $(V,\theta,\phi)$. So any Ansatz-based approach must work with functions of $3$ coordinates. Second, in the Reissner-Nordstr\"om case we have a dimensionless parameter $\e M$ that can be taken large to make gluing easier. In vacuum there is no analogue of this parameter. For this problem, it is likely to be beneficial to search for a low regularity solution first, even $C^0$. Indeed, as we observed in our results, many features of the $C^0$ solution persist in the higher regularity solutions. Furthermore, since all higher-regularity solutions must also be $C^0$ solutions, the absence of the latter would be strong evidence in favour of a third law for vacuum black holes.

\subsection*{Acknowledgements}
We are particularly grateful to Christoph~Kehle for his insightful discussions and suggestions, and to Carsten Gundlach for noticing errors in an earlier version of this paper. We also thank John~R.~V.~Crump and Ryan~Unger for helpful discussions. MG is supported by an STFC studentship and a Cambridge Trust Vice-Chancellor's Award. HSR and JES are supported by STFC grant ST/X000664/1, and JES is also supported by Hughes Hall College.

\appendix
\section{Numerical Implementation}\label{sec:numerics}

We have implemented our numerics in Mathematica 14. To begin, we input the equations of motion, Eqs.~\eqref{eq:EMCSF_spherical}, into the programme and use the symbolic differentiation features of the language to derive wave equations for higher-order quantities, such as $\partial_U^2r$ for $C^2$ gluing. To reduce computational costs, it is wise to attempt to solve the ODEs in one pass as a coupled system. Unfortunately, the process of deriving higher-order equations generates terms with lower-order derivatives, which at this stage have not yet been solved. To resolve this issue, we substitute the lower-order evolution equations back into the higher-order ones until we reach a coupled system of ODEs involving only the lowest-order transverse derivatives (e.g.~$\partial_Ur$, $\partial_U\Omega$, $\partial_U\Phi$, etc.).

In practice, instead of solving all of the equations in one pass, we first solve Raychaudhuri's equation, \eqref{eq:EMCSF_RayV}, backwards from $V=1$, as we know precisely which conditions to impose: $r(1)=r_+$ and $\partial_Vr(1)=0$. Then we solve the remaining functions forwards from $V=0$ in a single pass.

We perform the evolution with a standard 4th-order Runge-Kutta method. Most of our computations are done with a step-size of $0.001$, but we have regularly checked the accuracy of our results by conducting convergence tests with smaller step-sizes.

By far the most challenging aspect of the numerics is optimising the parameters $\alpha$ to obtain a candidate solution. We have opted to employ Broyden's ``good'' method \cite{Broyden:1965isd} to tackle this problem. This is a quasi-Newton root-finding method which, instead of explicitly computing the Jacobian, estimates it at each step. We initialise the Jacobian by computing it using finite difference methods.

Unfortunately, Broyden's method is not stable, and is not guaranteed to converge. Indeed, we find that unless the initial guess is quite close to a solution, then the method tends to fail. This is especially true in higher-dimensional spaces, i.e.~the higher the regularity class $C^k$ we are seeking a solution for, the more difficult it is to obtain said solution. It has sometimes been useful to scan over $\hat{\alpha}\in S^{2k}$, determine the value of $\beta$ such that $Q(1)=\q M$ with $\alpha=\beta\hat{\alpha}$, and sort by ascending order of the quantity $|(\partial_U\Phi,\dots,\partial_U^k\Phi)|$ to attempt to localise a solution to some subset of $S^{2k}$. Choosing an array of initial guesses in that subset and running Broyden's method would sometimes yield a solution.

Once a solution is found (candidate or valid) for some values of the parameters $(\q,\e M,\m/\e,\Lambda M^2)$, it is relatively easy to obtain solutions for nearby choices of parameters. This allows us to straightforwardly trace how the solution behaves as a function of these parameters. Our numerical data are available at \cite{data}.

\section{Extension of Kehle-Unger with Cosmological Constant}\label{sec:dur_cosmo_proof}

Kehle and Unger's construction of gluing data extends easily to the case with non-vanishing cosmological constant. First note that Lemmas 4.4 and 4.5 of \cite{Kehle:2022uvc} are in terms of $r_+$, so the same argument works with $\Lambda\neq 0$. Lemma 4.6 also involves $\e$ and $\q$. Note that for $\Lambda\neq 0$, $Q(1)=\q Q_{\rm max} \neq \q M$, but since $Q_{\rm max}\lesssim1.07M$ for all $\Lambda M^2$, the result holds with a small adjustment of constants. The first part of Lemma 4.7 is identical. With $\Lambda\neq0$ the initialisation of $\partial_Ur(0)$ is different, but we still obtain $\partial_Ur(0)<0$ via \eqref{eq:dUr(0)}, assuming (in the case of $\Lambda>0$) that the initial sphere is isometric to a sphere in the de Sitter static patch. Moreover, the extra term in \eqref{eq:EMCSF_waveR} does not change its parity, so Lemma 4.9 is also identical. Thus the only result that changes is Lemma 4.8, which ensures that $\sup\partial_Ur<0$.

To prove an extension of Lemma 4.8 we proceed in a similar way (see also \cite{Marin:2024rxt}). Since $r\geq r_+/2>0$, due to the previous lemmas, it suffices to show that $\sup r\partial_Ur<0$. First note that
\begin{equation}\label{eq:dv(rdur)}
    \partial_V(r\partial_Ur) = -\frac{\Omega^2}{4}\left[1-\frac{Q^2}{r^2} - \Lambda r^2\right].
\end{equation}
Clearly when $\Lambda<0$ this makes the RHS further negative than when $\Lambda=0$, and so we have $\sup\partial_Ur<0$ using the argument of \cite{Kehle:2022uvc}. Thus, we need only consider $\Lambda>0$. Choosing a gauge such that $\Omega=1$ on the horizon, we have
\begin{equation}
    |\partial_V(r\partial_Ur)| = \frac{1}{4}\left|1-\frac{Q^2}{r^2} - \Lambda r^2\right|.
\end{equation}
Now $Q(V)\leq Q(1) = \q Q_{\rm max} \leq c_0 r(V)$, where $c_0=c_0(\Lambda M^2)$ is a constant depending on $\Lambda M^2$ and we used $r\geq r_+/2$ from Lemma 4.4. This is the same step as in \cite{Kehle:2022uvc}. Moreover, $\Lambda M^2<2/9$ so $r^2\Lambda\leq 2r^2/9M^2 \leq 2r_+^2/9M^2\leq c_1$ for some constant $c_1$. (One can choose $c_1$ to be independent of $\Lambda M^2$.) If $1-Q^2/r^2-\Lambda r^2>0$ then we automatically have $\sup\partial_Ur<0$. So let us consider the opposite case. Since all terms are bounded we have
\begin{equation}
    |\partial_V(r\partial_Ur)| \leq \frac{1}{4}|1-c_0^2-c_1|\leq c_2
\end{equation}
for some constant $c_2=c_2(\Lambda M^2)$. A priori this constant depends on $\Lambda M^2$, but it may be that by considering the last two terms of \eqref{eq:dv(rdur)} together, one can choose $c_2$ independently of the cosmological constant (see the remark at the end). Integrating the previous expression, we have,
\begin{equation}
    \sup r(V)\partial_Ur(V) \leq r(0)\partial_Ur(0) + c_2.
\end{equation}
Estimations from previous lemmas \cite{Kehle:2022uvc} show that $\partial_Vr(0)\leq c_3\beta^2r_+\leq c_4\q/\e$ for constants $c_3$, $c_4=c_4(\Lambda M^2)$. Finally, using \eqref{eq:dUr(0)},
\begin{equation}
    -r(0)\partial_Ur(0) = \frac{r(0)[1-\Lambda r(0)^2/3]}{4\partial_Vr(0)} \geq \frac{\e}{4\q c_4}\frac{r_+}{2}[1-\Lambda r(0)^2/3]\geq c_5\frac{\e}{4\q}\frac{r_+}{2} \geq c_6 \frac{\e r_+^2}{\q Q_{\rm max}},
\end{equation}
where $c_5=c_5(\Lambda M^2)$, $c_6=c_6(\Lambda M^2)$ are constants, and where we again used that $1-\Lambda r(0)^2/3>0$ as we assume that the initial sphere is isometric to a sphere in the de Sitter static patch. Thus,
\begin{equation}
    \sup r(V)\partial_Ur(V) \leq -c_6 \frac{\e r_+^2}{\q Q_{\rm max}}+c_2
\end{equation}
and by taking $\e r_+^2/\q Q_{\rm max}$ sufficiently large we obtain $\sup\partial_Ur<0$. Note that in the original work, the quantity to be made large is $\e M/\q$. When $\Lambda=0$ this is identical to $\e r_+^2/\q Q_{\rm max}$ up to a constant.

Note that in the interval $\Lambda M^2\in[0,2/9]$, the constants $c_i$ can be set independently of $\Lambda M^2$ by choosing them to be large enough. However, it is not clear that this result extends to the full range of $\Lambda M^2\in(-\infty,2/9]$. For instance, as $\Lambda M^2\to-\infty$, $Q_{\rm max}/r_+$ diverges so $c_0$ must be arbitrarily large. Nevertheless, it may be possible to keep $c_2$ bounded, since the final term of \eqref{eq:dv(rdur)} is now positive, thereby reducing the effect of the second term. In practice, we find numerically that if $\e r_+^2/\q Q_{\rm max}$ is kept fixed then decreasing $\Lambda M^2$ improves the value of $\sup\partial_Ur$ of a valid solution.


\bibliographystyle{JHEP}
\bibliography{references.bib}

\end{document}